\documentclass[apjl]{emulateapj}
\usepackage{graphicx}
\usepackage{natbib}
\usepackage{amsmath}
\usepackage{multirow}
\usepackage{array}
\usepackage{subfigure}
\usepackage{float}

\begin{document}

\title{Mass-Radius Relations and Core-Envelope Decompositions of Super-Earths and Sub-Neptunes}
\author{Alex R. Howe}
\affil{Department of Astrophysical Sciences, Princeton University}
\affil{Peyton Hall, Princeton, NJ 08544, USA}
\affil{arhowe@astro.princeton.edu}
\author{Adam Burrows}
\affil{Department of Astrophysical Sciences, Princeton University}
\affil{Peyton Hall, Princeton, NJ 08544, USA}
\affil{burrows@astro.princeton.edu}
\author{Wesley Verne}
\affil{Department of Computer Science, Princeton University}
\affil{Princeton, NJ 08544, USA}


\begin{abstract}
Many exoplanets have been discovered with radii of 1$-$4 R$_\earth$, between that of Earth and Neptune. A number of these are known to have densities consistent with solid compositions, while others are ``sub-Neptunes'' likely to have significant H$_2$-He envelopes. Future surveys will no doubt significantly expand these populations. In order to understand how the measured masses and radii of such planets can inform their structures and compositions, we construct models both for solid layered planets and for planets with solid cores and gaseous envelopes, exploring a range of core masses, H$_2$-He envelope masses, and associated envelope entropies.  For planets in the super-Earth/sub-Neptune regime for which both radius and mass are measured, we estimate how each is partitioned into a solid core and gaseous envelope, associating a specific core mass and envelope mass with a given exoplanet. We perform this decomposition for both ``Earth-like'' rock-iron cores and pure ice cores, and find that the necessary gaseous envelope masses for this important sub-class of exoplanets must range very widely from zero to many Earth masses, even for a given core mass.  This result bears importantly on exoplanet formation and envelope evaporation processes.
\end{abstract}

\maketitle


\section{Introduction}
\label{intro}

\begin{figure}[htp]
\includegraphics[bb = 18 144 330 718,clip,width=\columnwidth]{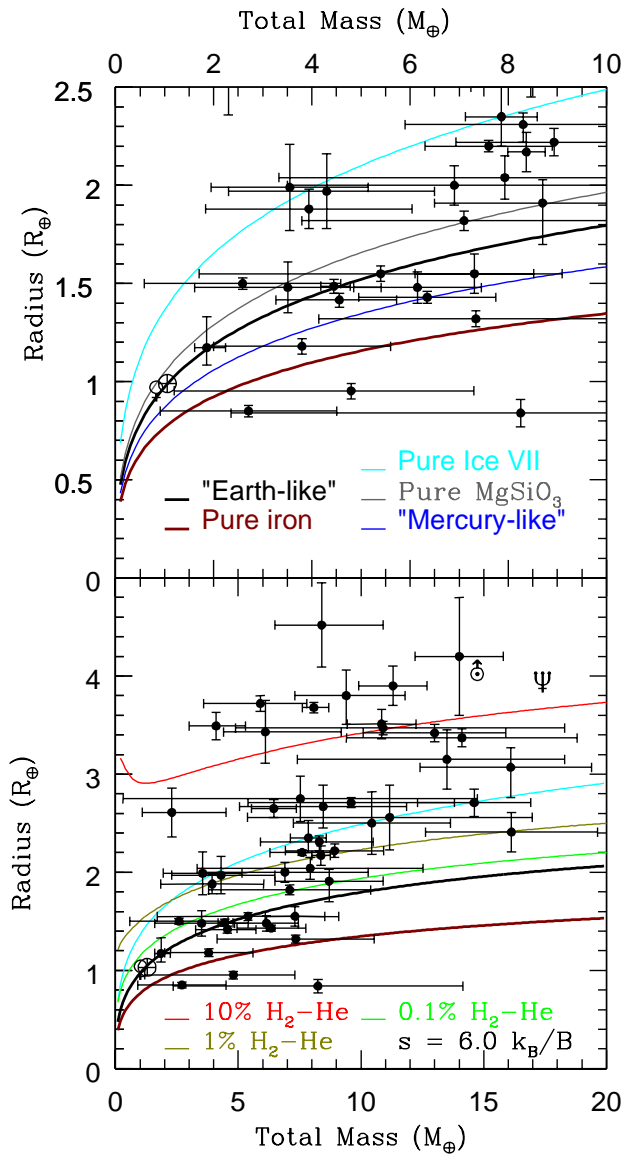}
\caption{Comparison of known exoplanets and Solar System planets with our models of simple Fe core/MgSiO\(_3\) mantle planets, pure water (Ice VII) planets, and planets with H$_2$-He envelopes. Earth-like is defined as 32.5\% core mass fraction (CMF) and Mercury-like is defined as 70\% CMF \citep{2007ApJ...669.1279S}. The range of possible mass/radius values for several planets lie squarely within the area occupied by Fe core/MgSiO\(_3\) mantle planets, making them excellent terrestrial exoplanet candidates. A number of others lie between the pure MgSiO$_3$ and pure Ice VII curves, making them, perhaps, candidates for ``water worlds,'' or for possessing small H$_2$-He envelopes. Another population of planets at a wide range of masses is consistent only with deeper H$_2$-He envelopes. The measured mass and radius values and references for the plotted planets are given in Table \ref{planet_list}.}
\label{exoplanets}
\end{figure}

The detection of thousands of candidate exoplanets with a wide range of masses and radii motivates the study of the general structure of	planetary bodies. While early detection methods heavily favored large planets with masses and radii near those of Jupiter, the recent trend has been toward lower masses and radii, some of which appear to be terrestrial, e.g. Kepler-10b, which was recently confirmed as a planet with radius \(1.416^{+0.033}_{-0.036}R_\earth\), mass \(4.56^{+1.17}_{-1.29}M_\earth\), and average density $8.8^{+2.1}_{-2.9}$ g cm\(^{-3}\) \citep{2011ApJ...729...27B}. For comparison, the average densities of the Earth and Venus are \(5.5\) g cm\(^{-3}\) and \(5.2\) g cm\(^{-3}\), respectively.

\citet{2013arXiv1311.0329L} have suggested that it is likely that planets larger than about 1.75 R$_\earth$ (based on mass-radius relations) have hydrogen/helium envelopes that contribute significantly to their radii.  In particular, they find that a planet's radius alone provides a first-order estimate of its composition, specifically, the H$_2$-He mass fraction. There is some uncertainty in this limit. For example, \citet{2014ApJ...783L...6W} adopt a maximum solid planet radius of 1.5 R$_\earth$, based in a maximum in the density distribution at $\sim$1.5 R$_\earth$ and $\sim$7.6 M$_\earth$ and \citet{Marcy.in.press} interpret this as a transition radius of 2.0 R$_\earth$, given an observed decrease in density from 1.5 to 2.0 R$_\earth$.

Similarly, \citet{2011ApJ...727...86R} suggests that envelope accretion onto a core, leading to a significant gaseous envelope, begins at a core mass of 10 M$_\earth$, or perhaps larger if the planets form close to their stars. However, recent observations of known exoplanets suggest that envelope accretion begins, on average, at a lower mass (as found by \citealt{2014ApJ...783L...6W}), and some individual planets appear to acquire gaseous envelopes at very low masses. For example, Kepler-51b has been measured to have a mass of $2.1_{-0.8}^{+1.5}$ M$_\earth$ and a radius of $7.1\pm 0.3$ R$_\earth$, corresponding to a density of 0.03$^{+0.02}_{-0.01}$ g cm$^{-3}$ \citep{2014arXiv1401.2885M}, clearly indicating a mostly gaseous composition.

Recent space-based missions such as {\it Kepler} \citep{2010Sci...327..977B} and {\it CoRoT} \citep{2006ESASP1306...33B} had photometric precision capable of measuring transits by Earth-sized planets. In the first 16 months of the {\it Kepler} Mission, 207 Earth-sized (\(R_p<1.25 R_\earth\)) and 680 super-Earth-sized (\(1.25 R_\earth < R_p < 2 R_\earth\)) planetary candidates were reported \citep{2013ApJS..204...24B}, suggesting a large number of solid planet candidates given a $\sim$1.75 R$_\earth$ cutoff. Figure \ref{exoplanets} provides a comparison of planets with measured radii and masses with theoretical mass-radius curves that we have generated for various simple planet compositions, including pure iron, Earth-like, Mercury-like, and pure silicate.\footnote{We define ``Earth-like'' to refer to a composition of 32.5\% Fe and 67.5\% MgSiO$_3$, and, when specifying a planet with a significant volatile mass fraction, a 13:27 ratio of Fe to MgSiO$_3$ mass. Similarly, we define ``Mercury-like'' to refer to a composition of 70\% Fe and 30\% MgSiO$_3$, and a 7:3 ratio if a significant volatile mass fraction is specified \citep{2007ApJ...669.1279S}. Note, however, that recent observations from MESSENGER suggest that Mercury's iron content may be closer to 73\% \citep{2013JGRE..118.1204H}.} We also include on Figure \ref{exoplanets} a pure water$-$in the form of Ice VII$-$mass-radius curve and three curves for models with gaseous envelopes: an ``Earth-like'' solid core and H$_2$-He mass fractions of 0.1\%, 1\%, and 10\%.\footnote{Throughout our paper, we use an envelope composition of 75\% H$_2$ and 25\% He by mass.}

\begin{figure}[htp]
\includegraphics[width=\columnwidth]{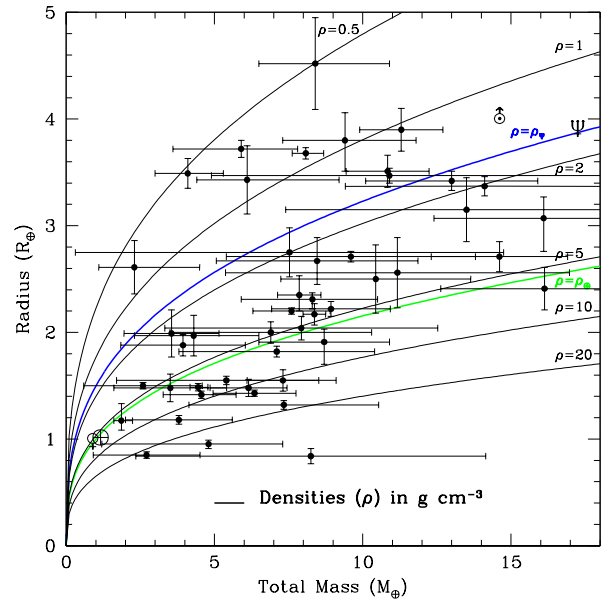}
\caption{Known extrasolar planets plotted against constant density curves, including Earth's density (green) and Neptune's density (blue). Densities ($\rho$) are given in g cm$^{-3}$.}
\label{planets_density}
\end{figure}

As Figure \ref{exoplanets} demonstrates, the mass/radius values of several known planets are consistent with an iron/rocky composition, and consistent with an Earth-like composition in particular.\footnote{One planet on the plots (Kepler-131c) has a measured density too high to be consistent with even a pure iron composition, and two others (Kepler-68c and Kepler-406c) have densities that are consistent only with the pure iron curves, likely due to the difficulty of making accurate mass measurements for smaller planets.} On the other hand, a number of observed planets have densities between those of pure water and pure silicate and are consistent with models of both ``water-worlds'' with a high water content and no significant envelopes, and models with H$_2$-He envelopes. Based on our models, the radius of a pure water, 10-M$_\earth$ planet is 2.5 R$_\earth$, which is near an observed break in the planet occurrence function in {\it Kepler} observations \citep{2012ApJS..201...15H,2013ApJ...778...53D}. However, it is not known which of these models (if either) is dominant in this radius range or whether this break reflects an actual difference in composition between planets smaller and larger than 2.5 R$_\earth$.

In Figure \ref{exoplanets}, we note a very wide spread in the mass-radius distribution for planets more massive than $\sim$2 M$_\earth$, with a variation of $\sim$2 R$_\earth$ at a given mass. For planets $\lesssim$8 M$_\earth$, this range overlaps with an Earth-like composition with no significant gaseous envelope. For higher-mass planets, some of these are also consistent with a no-envelope model if they have a sufficient water content, but we also observe planets with large radii that are consistent only with a structure that includes a deep H$_2$-He envelope, even at low masses ($\gtrsim$2 M$_\earth$).

In Figure \ref{planets_density}, we plot the extrasolar planets against constant density curves. Earth-density (green) and Neptune-density (blue) curves are included. The figure shows that density can vary by a factor of $\sim$5 between individual planets with the same radius. This large scatter makes it very difficult to fit any precise trends in radius with increasing mass.

In order to determine whether a given mass/radius pair indicates a solid composition, however, solid planet models and models of planets with gaseous envelopes must be constructed and compared with the data. These exoplanets may potentially have a wide range of possible compositions and temperatures and a similarly wide range of possible gaseous envelopes, so models must be able to be adjusted accordingly. Solid exoplanet models are important in both cases, since they may be used to model solid cores of planets with gaseous envelopes by applying a non-zero-pressure boundary condition at the core-envelope interface.

While there is a rich history of exoplanet structural modeling, there are a number of important areas which have yet to be investigated. For planets with gaseous envelopes, the effects of irradiation and atmospheric heating are poorly understood, and the degeneracies of envelope mass, envelope entropy, and core mass have not been explored in detail. Moreover, the implications of the large scatter in the mass-radius distribution, particularly on the search for Earth-like planets, are only beginning to be addressed.

For solid planets, equations of state for planetary materials at the pressures found in planets are subject to a degree of uncertainty, and different EOS models produce different results, which warrant analysis of the uncertainty in the modeled mass and radius values.

Our paper investigates the uncertainties and degeneracies in exoplanet modeling, particularly of planets with H$_2$-He envelopes, in order to gain a better understanding of what is measurable in observed exoplanets. We compute mass-radius curves over the range of 0.1 to 20 M$_\earth$ for a variety of planet attributes, and explore the possibility of determining a precise core-envelope decomposition from mass and radius observations. We study planets with both ``Earth-like'' cores and ice cores, which may both be of interest depending on whether planets with gaseous envelopes form beyond the snow line. We also compute mass-radius curves for various compositional profiles for solid planets. Note that observations suggest that planets more massive than $\sim 4-8$ M$_\earth$ are likely to possess significant H$_2$-He envelopes \citep{2014ApJ...783L...6W}, and the same is expected to be true of planets with radii larger than 1.5$-$2.0 R$_\earth$ \citep{2013arXiv1311.0329L,2014ApJ...783L...6W,Marcy.in.press}, so our results for purely solid planets likely apply only to smaller and less massive objects. Conversely, our results for planets with H$_2$-He envelopes will apply to planets larger than $\sim 4-8$ M$_\earth$ or 1.5$-$2.0 R$_\earth$.

While e.g. \citep{2013arXiv1311.0329L} perform evolutionary calculations to produce planetary structural models, because of the large uncertainties in the parameters that go into these calculations, we do not do this, and we believe that they do not constitute an improvement over our method. Specifically, the ages of known exoplanets are, in most cases, not measured, so that a wide range of ages is possible. Metallicities affect opacities, which in turn affect the cooling rate. They also affect the mean molecular weight and scale height of the atomsphere. Because the formation mechanism of these planets is not known, it is not appropriate a priori to estimate their uncertainties by analogy with Uranus and Neptune. In addition, irradiation is usually treated in an ad hoc manner by averaging the stellar flux over the entire planet rather than modeling day-to-night heat redistribution, which \citet{2010ApJ...722..871S} show has a significant effect on the net cooling rate.

All of these effects will influence the outcome of an evolutionary model, resulting in large uncertainties. Therefore, we sidestep these uncertainties by focusing on a single variable, the entropy of the envelope, and compute structural models over a range of entropies. The entropy and surface gravity of a planet completely determine its structure (with a small correction for metallicity), so our models are, in effect, the same as those provided by evolutionary calculations.

Section \ref{previous} summarizes the previous work modeling solid exoplanets and sub-Neptunes. Section \ref{hydrogen} presents our models with H$_2$-He envelopes and the effect on radius of varying the core mass, envelope mass, and entropy in the envelope. In Section \ref{fits}, we explore the core-envelope decomposition and produce a fit to the mass-radius relation of \citet{2014ApJ...783L...6W}. Section \ref{error} gives an overview of the quantitative effects of varying our model parameters. Section \ref{internal} demonstrates our code's output with density-pressure profiles of planets, central pressures and densities, and envelope base pressures. Section \ref{solid} presents our models of solid planets, and Section \ref{conclusions} summarizes our overall conclusions. Our modeling procedure, associated code, and our studies and selection of our equations of state are described in the Appendix.


\section{Previous Work on Solid Exoplanets and Sub-Neptunes}
\label{previous}

Early efforts to calculate mass-radius relationships for planetary bodies of various compositions were made by \citet{1969ApJ...158..809Z}, using a Thomas-Fermi-Dirac equation of state (TFD EOS) described in \citet{1967PhRv..158..876S}. Those authors integrated the equations of hydrostatic equilibrium in conjunction with their EOS and a zero-pressure surface boundary condition to construct planetary structural models. This is the standard procedure for solid planet modeling; most recent advances have been in the accuracy of the low-pressure EOS, driven by the availability of experimental pressure/density data \citep{2001GeoRL..28..399A}.

The TFD EOS is valid only in the high-pressure limit where electrons are a non-interacting degenerate gas, but \citet{1967PhRv..158..876S} used a correlation energy correction to account for interactions between the electrons at lower pressures. They thereby extended the validity of their EOS down to \(\sim\)1 Mbar (by their estimation). However, as the TFD EOS does not account for chemical structure, it has zero-pressure density errors up to a factor of 2 \citep{1969ApJ...158..809Z}. As such, \citet{1969ApJ...158..809Z} focused on high-mass planets whose internal pressures lay largely in the $\gtrsim$1 Mbar regime, calculating a ``critical mass'' for various compositions beyond which a planet's radius decreases with additional mass. They investigated only simple monatomic elemental compositions (pure H, He, C, Mg, Fe, and various H/He mixtures) because these are most easily modeled by the TFD EOS, which considers each element separately \citep{1967PhRv..158..876S}. In addition, \citet{1969ApJ...158..809Z} derived the maximum radius, and the mass and central pressure at which the maximum radius is achieved, as a function of He ratio in a H$_2$-He planet. Their models assumed a constant composition throughout the planet with no core-mantle-envelope differentiation, which limits their applicability for solid exoplanets.

More recent work implements equations of state based on experimental data. An early example of this approach is \citet{1982AREPS..10..257S}, who used contemporary shock wave data as the basis for his low-pressure equations of state for ices (H\(_2\)O, CH\(_4\), and NH\(_3\)). \citet{1982AREPS..10..257S} also investigated the interior structure and composition of giant planets, and produced mass-radius diagrams for various compositions. He reported a lack of accurate equations of state available for ferromagnesian rock, and as such, solid planet models were outside of the scope of his paper.

In the last decade, a number of authors have presented models of solid exoplanets, motivated by the aforementioned exoplanet detections, as well as the increased availability of valid semi-empirical models for terrestrial materials. \citet{2006Icar..181..545V} defined and modeled two exoplanet classes: ``super-Earths,'' with similar compositions to Earth and planet mass 1 M$_\earth < M_p < 10$ M$_\earth$, and ``super-Mercuries,'' with similar compositions to Mercury and planet mass 1 M$_{\textrm{Mercury}} < M_p < 10$ M$_{\textrm{Mercury}}$. It should be noted, however, that the term ``super-Earth'' is now often used to refer to any ``terrestrial'' planet with a mass greater than that of Earth, as well as planets in the 1$-$10 M$_\earth$ range \citep{2011arXiv1108.0031H}, or the 1.25$-$2.0 R$_\earth$ range \citep{2013ApJS..204...24B}. The models of \citet{2006Icar..181..545V}, along with many other contemporary models \citep{2007Icar..191..337S,2007ApJ...669.1279S,2007ApJ...659.1661F}, used a fourth-order Runge-Kutta integration scheme to solve the equations of hydrostatic equilibrium. The equations of state used by \citet{2006Icar..181..545V} were zero-temperature Birch-Murnaghan (B-M) equations of state \citep{2000ipei.book.....P} with thermal corrections using a Debye model. The B-M EOS is based on low-temperature pressure/density data. Because there are limits to the pressures that such experiments can reach, the B-M EOS incorporates an extrapolation to higher pressures. Though the thermal corrections to the equations of state for rocky materials are generally small, the model of \citet{2006Icar..181..545V} required a detailed temperature profile in order to calculate the phase transitions in the silicate mantle. Their thermal model relies on the assumption of convective heat transport in the core and mantle, with conductive layers at the core-mantle boundary and the surface. They iterated their model, using the compositional profile to determine parameters to compute the temperature profile, which was used in turn to determine phase transitions for the compositional profile calculation, until a self-consistent planet model was achieved.

\begin{table*}[tbhp]
\caption{Equations of State Used in Recent Super-Earth Modeling Papers}
\begin{center}
\begin{tabular}{l|>{\raggedright\arraybackslash}m{4cm}|m{5cm}|m{1.85cm}}
\hline
Authors & Material & EOS & References \\ \hline \hline
\multirow{6}{*}{\citet{2006Icar..181..545V}} & 
\mbox{Fe; FeO; Fe+alloy;} 
\mbox{(Mg\(_{1-x}\), Fe\(_x\))\(_2\)SiO\(_4\)} \mbox{(olivine, wadsleyite,} ringwoodite); 
\mbox{(Mg\(_{1-x}\); 
Fe\(_x\))SiO\(_3\)} perovskite; 
\mbox{(Mg\(_{1-x}\), Fe\(_x\))O} & 3\(^\textrm{rd}\) order B-M, with Debye correction & 1,2,3,4 \\ \hline
\multirow{3}{*}{\citet{2007ApJ...656..545V}} & 
Same as \citet{2006Icar..181..545V}, plus H\(_2\)O (ice) & Vinet, with Debye correction & 2,3,5,6,7,8 \\ \cline{2-4}
& H\(_2\)O (liquid) & Rankine-Hugoniot & 9 \\ \hline
\multirow{3}{*}{\citet{2007ApJ...659.1661F}} & H\(_2\)O, olivine & ANEOS & 10 \\ \cline{2-4}
& iron & SESAME 2140 & 11 \\ \cline{2-4}
& H$_2$-He & \citet{1995ApJS...99..713S} & 12 \\ \hline
\multirow{3}{*}{\citet{2007Icar..191..337S}} & Same as \citet{2007ApJ...656..545V} (\mbox{but different} \mbox{H\(_2\)O [liquid] EOS}) & 3\(^\textrm{rd}\) order B-M & 2,6,13,14, 15,16,17,18 \\ \cline{2-4}
{\citet{2009ApJ...693..722G}} & H\(_2\)O (liquid) & 2\(^\textrm{nd}\) order B-M & 19 \\ \hline
\multirow{10}{*}{\citet{2007ApJ...669.1279S}} & \mbox{C (graphite); Fe (\(\alpha\))}; \mbox{FeS; H\(_2\)O (ice VII)}; MgO; \mbox{MgSiO\(_3\) (enstatite)}; \mbox{[Mg,Fe]SiO\(_3\) (perovskite);} SiC & 3\(^\textrm{rd}\) order B-M & 5,15,20,21, 22,23,24, 25,26 \\ \cline{2-4}
& Fe (\(\epsilon\)) & Vinet & 27 \\ \cline{2-4}
& H\(_2\)O (liquid) & logarithmic EOS & 28 \\ \cline{2-4}
& H\(_2\)O (VII$-$X transition) & tabular DFT calculations & 29 \\ \cline{2-4}
& All (high pressure) & Thomas-Fermi-Dirac (TFD) & 30 \\ \hline
\multirow{3}{*}{\citet{2011ApJ...738...59R}} & Same as \citet{2006Icar..181..545V} & Same as \citet{2006Icar..181..545V} & 5,15,20-30 \\ \cline{2-4}
& H$_2$-He & \citet{1995ApJS...99..713S} & 12 \\ \hline
\end{tabular}
\end{center}
\tablerefs{
(1) \citet{2003JGRB..108.2045L};
(2) \citet{2001JGR...10621799U};
(3) \citet{1997PEPI..100...49W};
(4) \citet{Anderson_Isaak_2000};
(5) \citet{1987Natur.330..737H};
(6) \citet{1993JChPh..99.5369F};
(7) \citet{2005GeoJI.162..610S};
(8) \citet{2004EaPSL.224..241T};
(9) \citet{2005JGRE..11003005S}, gives constraints based on shock data;
(10) \citet{Thompson1990};
(11) \citet{LyonJohnson1992};
(12) \citet{1995ApJS...99..713S};
(13) \citet{2001E&PSL.185...25K};
(14) \citet{Hemley1992};
(15) \citet{1995Natur.378..170D};
(16) \citet{1996GeoRL..23.1143B};
(17) \citet{1998PEPI..106..275V};
(18) \citet{1991JGR....9618037A};
(19) \citet{Lide2005};
(20) \citet{Ahrens2000};
(21) \citet{1989PhRvB..3912598H};
(22) \citet{1989PhRvB..40..993Z};
(23) \citet{King:a21332};
(24) \citet{Olinger1977};
(25) \citet{1987Sci...235..668K};
(26) \citet{1989JETPL..50..127A};
(27) \citet{2001GeoRL..28..399A};
(28) \citet{2003fuph.book.....H};
(29) Density functional theory calculations by \citet{2007ApJ...669.1279S};
(30) \citet{1967PhRv..158..876S}.
}
\label{eos_author_table}
\end{table*}

\citet{2007ApJ...656..545V} applied this model to the exoplanet GJ 876d and introduced a water layer consisting of high-pressure ices covered by a thin liquid water ocean. They also used a Vinet EOS fit \citep{1989JPCM....1.1941V}, as opposed to the B-M EOS used by \citet{2006Icar..181..545V}, because the Vinet fit is reported to extrapolate better to high pressures \citep{1996JPCM....8...67H}. The model from \citet{2006Icar..181..545V} was also applied in \citet{2007ApJ...665.1413V} to investigate degeneracies in the iron core, silicate mantle, and H\(_2\)O mass fractions for a given planet mass and radius and to construct ternary diagrams showing curves of constant radius for a given mass.

\citet{2007Icar..191..337S} used a similar physical approach to that of \citet{2006Icar..181..545V}, but employed the stellar composition (minus H$_2$ and He) of the planet in constructing structural models. This approach is justified by observations that meteorite chemical ratios (thought to be representative of early planets) are similar to those found in the Sun. They used five independent parameters to determine the composition and internal structure of the planet: Mg/Si, Fe/Si, Mg\# (defined as the mole fraction Mg/(Mg + Fe) in silicates), H\(_2\)O mass fraction, and total mass. They also determined a mass-radius model for planets with a water ocean.

\citet{2007ApJ...669.1279S} conducted a broader investigation of solid exoplanets by using a simpler zero-temperature model that incorporated the Thomas-Fermi-Dirac EOS at high pressures with the Vinet semi-empirical EOS at lower pressures. They did not address phase transitions in the silicate mantle because phase transitions have little effect on the mass-radius curve of a given material and require a temperature profile. Instead, they assumed a constant-composition MgSiO\(_3\) (perovskite) mantle\footnote{In our paper, (perovskite) or (olivine), placed after a chemical formula, refers to the crystal structure, not the specific compound or precise chemical make-up.}. These simplifications allowed them to investigate a wide range of planet compositions and masses.

\citet{2007ApJ...659.1661F} took an even broader approach, investigating five orders of magnitude in mass (0.01 Earth masses to 10 Jupiter masses) and a variety of planetary compositions, as well as envelopes. For the solid components of their planets, they used a model similar to that of \citet{2007ApJ...669.1279S}, though they used Mg\(_2\)SiO\(_4\) (olivine) for the mantle instead of MgSiO\(_3\) (perovskite), and used tabular EOS data from the ANEOS \citep{Thompson1990} and SESAME \citep{LyonJohnson1992} compilations, as opposed to semi-empirical fits. They neglected thermal corrections for the Mg\(_2\)SiO\(_4\) (olivine) and iron equations of state, but for water they used a thermal EOS correction of the form
\begin{equation}
P=P_0+3.59\times 10^{-5}\rho T,
\end{equation}
where \(P\) is the corrected pressure in Mbar, \(P_0\) is the zero-temperature pressure in Mbar, \(\rho\) is the density in g cm\(^{-3}\), and \(T\) is the temperature in Kelvin. Their main goal was to produce a general, if very approximate, theory for comparison with observational data, and as such they neglected the details found in some previous papers.

\citet{2009ApJ...693..722G} extended the work of \citet{2007Icar..191..337S} to masses of 100 M$_\earth$ and also compared it with contemporary models to determine how precisely planetary compositions can be determined from mass and radius data, in particular, the water mass fraction. They found that, given uncertainties in internal structure, the water fraction can be determined with a standard deviation of 4.5\% if the mass and radius are known exactly, but this uncertainty increases rapidly with the uncertainty in the radius.

\citet{2011ApJ...738...59R} also investigated planets with significant H$_2$-He envelopes in the context of estimating plausible masses of {\it Kepler} planet candidates of radius 2$-$6 R$_\earth$. They considered planet models with up to four layers: iron, silicates, water, and a hydrogen-helium envelope. They defined the exterior boundary condition as the radius at which the radial optical depth of the atmosphere is $\tau_R = 2/3$. They used the same EOS as \citet{2007ApJ...669.1279S} for the solid components and the tabular EOS of \citet{1995ApJS...99..713S} for the gaseous envelopes. They computed a temperature profile based on radiative transfer and radiative diffusion in the outer part of the envelope, transitioning to an adiabatic profile at the onset of convection. They considered planets produced by simulations of both core-nucleated accretion and outgassing of volatiles, particularly hydrogen. In the case of core-nucleated accretion, they considered cores of 10\% iron, 23\% silicates (Fe$_{0.1}$Mg$_{0.9}$SiO$_3$), and 67\% water and solar-composition envelopes. They computed mass-radius curves for models with envelope mass fractions from 0.001 to 0.5, characteristic specific powers from $10^{-12.5}$ to $10^{-9.5}$ W kg$^{-1}$, and equilibrium temperatures of 500 K and 1000 K. For outgassing-produced envelopes, they modeled the reaction of water with iron (which produces a pure hydrogen atmosphere) on planets with water mass fractions from 8.6\% to 20\%.

A summary of the equations of state used in previous models of solid exoplanet structure is given in Table \ref{eos_author_table}.


\section{Planets With H$_2$-H\MakeLowercase{e} Envelopes}
\label{hydrogen}

\begin{figure}[htp]
\includegraphics[bb = 18 144 330 718,clip,width=\columnwidth]{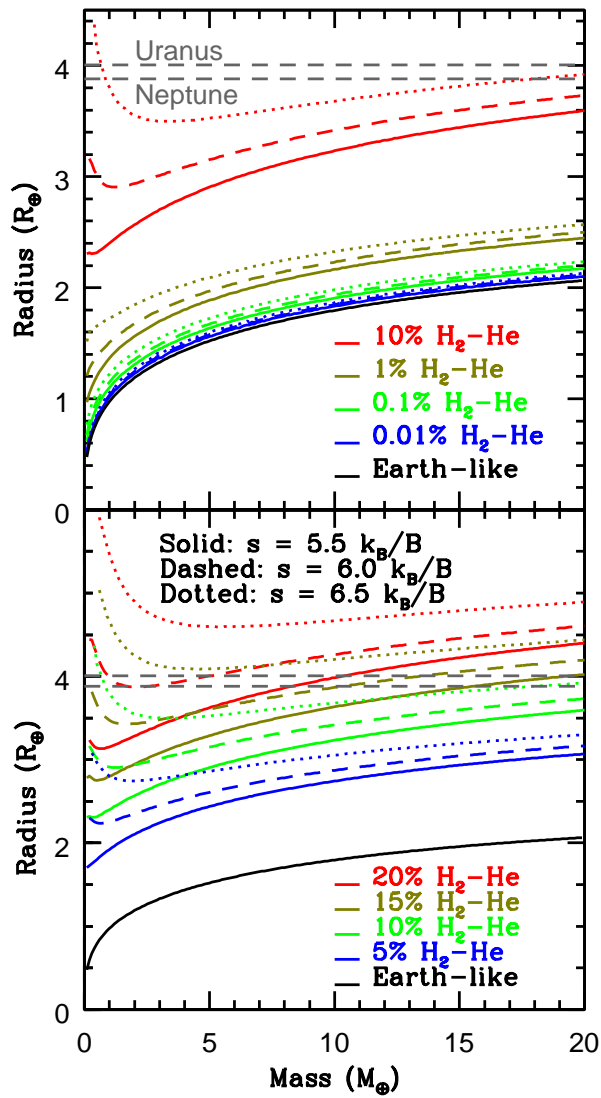}
\caption{Mass-radius curves of planets with Earth-like cores and gaseous envelopes. Top panel: envelopes equal to 0.01\%, 0.1\%, 1\%, and 10\% of the total mass. Bottom panel: envelopes equal to 5\%, 10\%, 15\%, and 20\% of the total mass. Curves with envelope entropies of 5.5, 6.0, and 6.5 $k_B$ per baryon ($k_B/B$) are plotted (assuming a convective envelope). An upturn in radius at low mass is apparent for larger envelope fractions. In the most extreme case of 20\% H$_2$-He and $s = 6.5 k_B/B$, the minimum radius occurs at a mass of 5.5 M$_\earth$. The radii are very sensitive to the H$_2$-He fraction, but much less sensitive to the total mass of the planet, particularly for masses $>5$ M$_\earth$.}
\label{varmass}
\end{figure}

We now present new models of planets with H$_2$-He envelopes with both ``Earth-like'' rock-iron cores with 32.5\% Fe and 67.5\% MgSiO$_3$ (perovskite) and with pure water (Ice VII) cores. Ice cores are of particular interest because it may be the case that planets with significant gaseous envelopes and core masses form only beyond the snow line. We compute models with varying envelope masses from 0 to 10 M$_\earth$ and envelope entropies ranging from 5.5 to 6.5 $k_B$ per baryon ($k_B/B$)$-$in most cases, using discrete values of 5.5, 6.0, and 6.5 $k_B$ per baryon. \footnote{We assume a convective envelope with constant entropy throughout. This entropy results from cooling and is a function of, among other things, age and metallicity and provides a convenient way to parameterize the uncertainties in these parameters, which would present serious problems in the case of evolutionary calculations.} These values are comparable to the entropies found by the evolutionary models of \citet{2013arXiv1311.0329L} for planets of Gyr age or older (specifically, Solar-metallicity models fall entirely within this range for $t>1$ Gyr, and 50x Solar enhanced-opacity models fall entirely within this range for $t>4$ Gyr).

In this section, we present models with a fully convective envelope, i.e., models with only a thin radiative atmosphere. This is a good approximation to Uranus and Neptune, where the radiative-convective boundary is at $<1$ bar \citep{2013arXiv1312.3323S}, and the equilibrium temperature is $\sim 50$ K, resulting in a small scale height. However, it is not a good approximation for highly-irradiated planets, for which the radiative-convective boundary is at a high pressure of $\sim 1000$ bar \citep{2013arXiv1312.3323S} and the equilibrium temperature is much larger. The depth of the radiative atmosphere varies widely depending on the irradtion level and surface gravity. We investigate the effect of this radiative atmosphere on computed masses and radii in Section \ref{fits}.

In Figure \ref{varmass}, we plot radius versus total mass for planets with Earth-like cores and H$_2$-He envelope with mass fractions ranging from 0.01\% to 20\%. The code produces results consistent with the known properties of Uranus and Neptune ($\sim$10\% H$_2$-He) and also reproduces the upturn in radius at low masses for envelopes comprising $\ge 5\%$ of the total mass, which was produced by \citet{2011ApJ...738...59R}.\footnote{For these compositions, the mass corresponding to the minimum radius on the mass-radius curves increases with H$_2$-He fractions and increases even faster with entropy, rising from $<0.1$ M$_\earth$ for 5\% H$_2$-He and $s = 5.5\,k_B/B$ (if, indeed, there is a local minimum) to 5.5 M$_\earth$ for 20\% H$_2$-He and $s = 6.5\,k_B/B$, among the models we study.} Figure \ref{varmass} also shows that the mass-radius curves are only slowly-rising for total masses $\gtrsim$5 M$_\earth$, i.e. radius is not strongly dependent on total mass, while it is more strongly dependent on the mass fraction and entropy in the H$_2$-He envelope.

\begin{figure}[htp]
\includegraphics[width=\columnwidth]{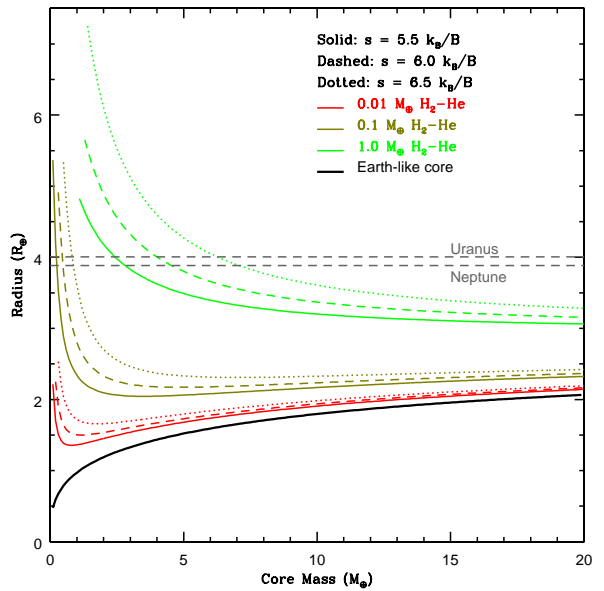}
\caption{Radius versus core mass for planets with constant envelope masses of 0.01, 0.1, and 1.0 M$_\earth$. Curves with envelope entropies of 5.5, 6.0, and 6.5 $k_B/B$ are plotted. The curves are remarkably flat for core masses greater than $\sim$5 M$_\earth$, indicating that the properties of the core have little influence on the observable properties of all but the smallest known planets.}
\label{mass_radius_varcore}
\end{figure}

Alternatively, in Figure \ref{mass_radius_varcore}, we plot radius versus {\it core} mass, rather than {\it total} mass, for constant envelope masses of 0.01, 0.1, and 1.0 M$_\earth$. Here, we see that for lower-mass cores ($\lesssim$5 M$_\earth$), the planetary radius is quite sensitive to core mass and entropy, but for higher-mass cores ($\gtrsim$5 M$_\earth$), which cover most current planet observations, the radius is most sensitive to envelope mass alone and varies very little with core mass, even less than with total mass. This suggests that mass-radius observations can be used to determine the core-envelope decomposition for a planet more precisely than the envelope mass fraction. In particular, because mass measurements usually have much larger uncertainties than radius measurements, it will be possible in many cases to determine envelope mass with more precision than mass fraction, which will have useful applications in formation models. In Table \ref{0.1menv_6.0_table}, we provide a sample table of properties of these models as a function of $M_c$ for $M_{env}=0.1$ M$_\earth$ and $s=6.0 k_B$ per baryon.

\begin{figure}[htp]
\includegraphics[bb = 18 144 330 718,clip,width=\columnwidth]{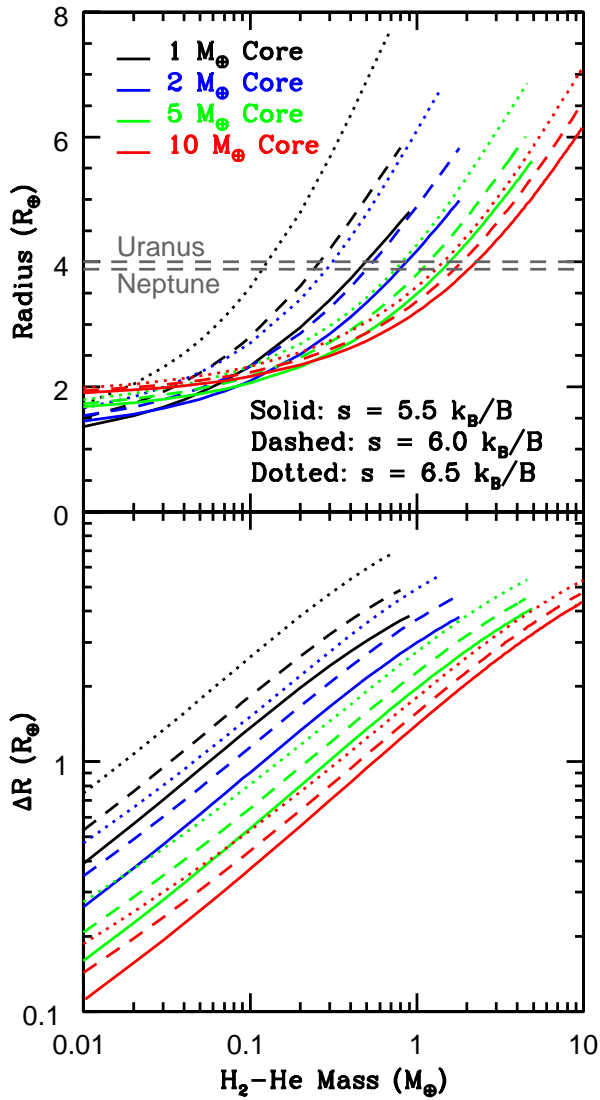}
\caption{Top panel: radius versus envelope mass for planets with constant core masses of 1, 2, 5, and 10 M$_\earth$. Bottom panel: Envelope depth ($\Delta R = R_p - R_c$) versus envelope mass. Curves with entropies of 5.5, 6.0, and 6.5 $k_B/B$ are plotted. Envelope depth follows a power law in terms of envelope mass, $\Delta R \propto M_{env}^x$, where $x=0.523-0.577$.}
\label{mass_radius_varenv}
\end{figure}

For comparison, in the top panel of Figure \ref{mass_radius_varenv}, we plot radius versus {\it envelope} mass for models with constant core masses of 1, 2, 5, and 10 M$_\earth$. Here, again, we see that for lower-mass cores (1 and 2 M$_\earth$), the planetary radius is quite sensitive to core mass and envelope entropy, but, for higher-mass cores (5 and 10 M$_\earth$), radius is most sensitive to envelope mass.

We also note that the curves are relatively flat for envelopes with masses $\lesssim$0.1 M$_\earth$, in which case the radius of the solid core dominates the total radius. However, we see another useful relation in the bottom panel of Figure \ref{mass_radius_varenv}, where we plot the envelope depth, $\Delta R = R_p - R_c$, versus envelope mass. In all cases, the envelope depth follows an approximate power law with core mass and envelope mass: $\Delta R \propto M_{env}^xM_c^y$. While the curves are bent to a slightly shallower slope at both low and high masses, the power-law indices over most of their lengths fall within a narrow range for $x$, $x=0.523\rightarrow 0.577$, but a wider range for $y$, $y=-0.565\rightarrow -0.693$. We provide tables of the properties of representative models from this plot in Tables \ref{2mc_5.5_table}-\ref{10mc_6.0_table}.

\begin{figure}[htp]
\includegraphics[width=\columnwidth]{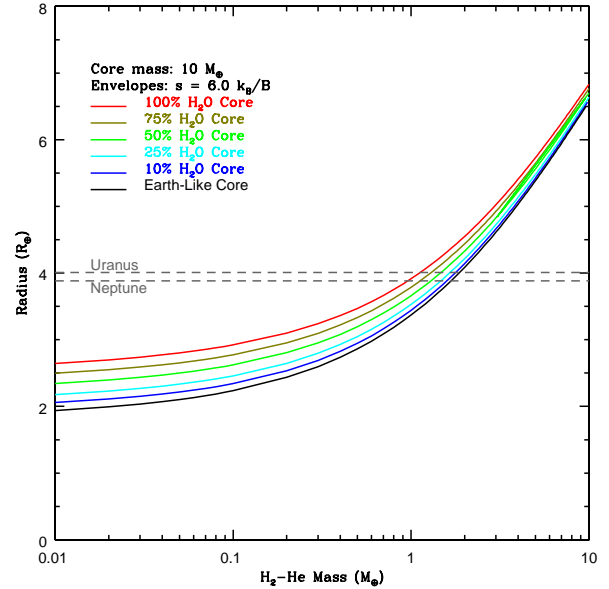}
\caption{Radius versus envelope mass for planets with 10-M$_\earth$ cores containing water layers (in the form of Ice VII) with mass fractions of 0\%, 10\%, 25\%, 50\%, 75\%, and 100\%. Envelope entropy is set to 6.0 $k_B/B$.}
\label{mass_radius_h2h2o}
\end{figure}

In Figure \ref{mass_radius_h2h2o}, we plot radius versus envelope mass for models with a constant core mass, 10 M$_\earth$, and entropy, $s=6.0\,k_B$ per baryon, but varying water fraction in the core$-$using a core structure with iron and silicate layers surrounded by a water (Ice VII) layer, comprision a varying fraction of the core mass from no water content to a pure ice core. The effect of the water fraction on radius is dominant in planets with small gaseous envelopes of $\lesssim 1$ M$_\earth$. For larger envelopes, the envelope mass becomes more important, and the variation with water fraction shrinks. Changing the water abundance of the core causes the total radius of the planet to vary by up to about 30\% if the envelope mass is small (so the effect of the change in the core radius is strong), but by only about 10\% for larger envelopes, $M_{env}\rightarrow 10$ M$_\earth$, less than the effect of varying the entropy for envelopes of these masses.

\begin{figure}[htp]
\includegraphics[width=\columnwidth]{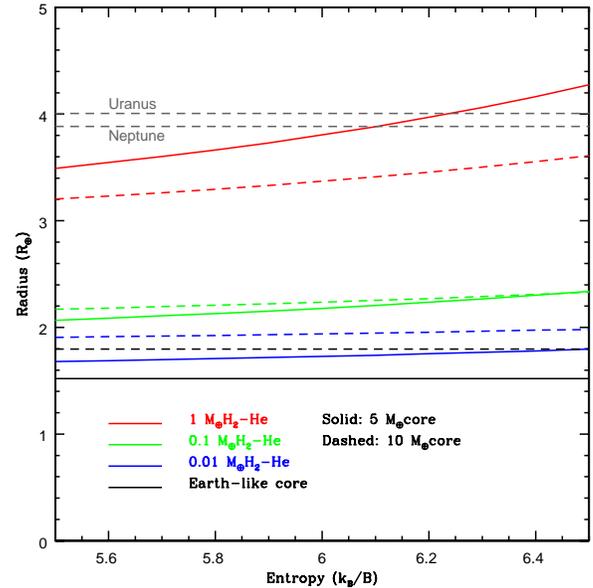}
\caption{Radius versus entropy for planets with core masses of 5 and 10 M$_\earth$. Curves with envelope masses of 0.01, 0.1, and 1.0 M$_\earth$ are plotted. Entropy becomes a significant influence on the radius for large envelopes.}
\label{mass_radius_vars}
\end{figure}

In Figure \ref{mass_radius_vars}, we set constant core masses of 5 and 10 M$_\earth$ and plot radius versus entropy for constant envelope masses of 0.01, 0.1, and 1.0 M$_\earth$. The radius is relatively insensitive to entropy for the lower envelope masses, but entropy becomes significant for envelopes with masses $\gtrsim$1 M$_\earth$.


\section{Fits to Known Planets}
\label{fits}

\begin{figure}[htp]
\includegraphics[width=\columnwidth]{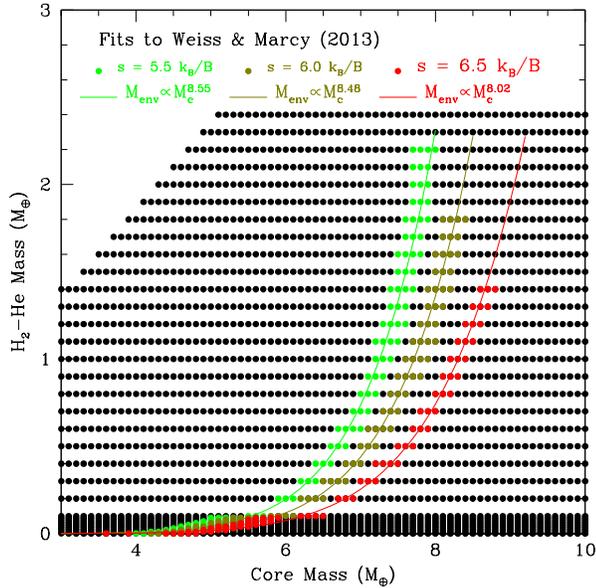}
\caption{The functional mass-radius fit of \citet{2014ApJ...783L...6W} plotted through a data cube of our planet models (points) with varying entropy, core mass, and envelope mass, for Earth-like (iron-silicate) cores. Colored points lie within 0.2 M$_\earth$ of this functional fit for an envelope entropy of 5.5 (green), 6.0 (yellow), and 6.5 (red) $k_B/B$. We fit a power law (shown) in (core mass)-(envelope mass) space to the functional fit for each entropy.}
\label{planets_grid}
\end{figure}

\begin{figure}[htp]
\includegraphics[width=\columnwidth]{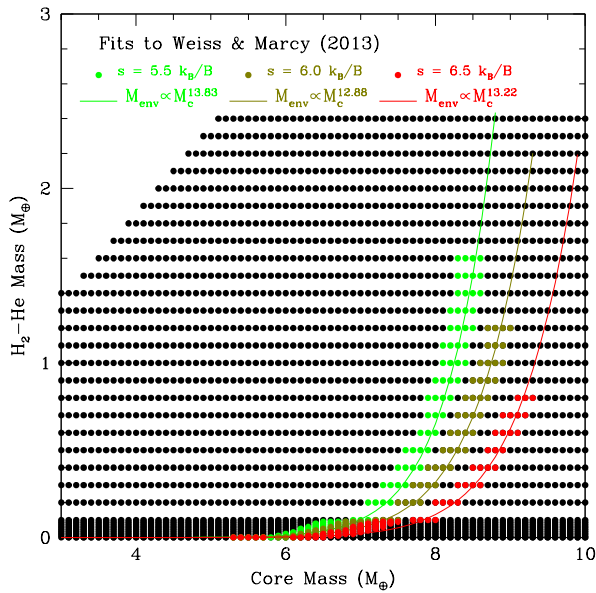}
\caption{Same as Figure \ref{planets_grid}, but for pure ice cores.}
\label{planets_h2o_grid}
\end{figure}

\citet{2014ApJ...783L...6W} fit a two-piece mass-radius function to 65 low-mass planets with radii $\le 4$ R$_\earth$ and masses measured by \citet{2014ApJS..210...20M}. For planets with radii $<1.5$ R$_\earth$, they fit an Earth-like model defined by the density formula $\rho = 2.43+3.39R$, where $\rho$ is the density in g cm$^{-3}$ and $R$ is the radius in Earth radii. For planets with radii $>1.5$ R$_\earth$, they apply a model of increasing H$_2$-He fraction with mass with a (nearly-linear) power law fit: $M = 2.69R^{0.93}$, where $M$ and $R$ are given in Earth masses and radii.

We fit these mass-radius fits to a data cube of our models with varying entropy, core mass, and envelope mass in Figures \ref{planets_grid} and \ref{planets_h2o_grid}. Ambiguities in metallicity, age, and heat redistribution make it difficult to investigate the exact structures of individual planets, so we seek to bracket the range of possibilities with different core compositions and entropies. In Figure \ref{planets_grid} we employ models with Earth-like cores (32.5\% Fe and 67.5\% MgSiO$_3$), and in Figure \ref{planets_h2o_grid}, we employ models with pure water (Ice VII) cores. We plot each of our models as a point in (core mass)-(envelope mass) space. Each point corresponds to a particular total mass and a range of radii, depending on the envelope entropy. By setting the entropy to 5.5 (green), 6.0 (yellow), and 6.5 (red) $k_B$ per baryon, we highlight those points that lie within 0.2 M$_\earth$ of the mass-radius fit derived by \citet{2014ApJ...783L...6W}. Following these highlighted points, we fit power laws in (core mass)-(envelope mass) space to the functional fit at each entropy. In general, we find a good fit to a power law, $M_{env} \propto M_c^x$, where $x = 8.0-8.5$ for rock-iron cores and $x = 13-14$ for pure ice cores, a very steep dependence of envelope mass on core mass.

\begin{figure*}[htp]			
\begin{center}
\subfigure{\includegraphics[width=0.49\textwidth]{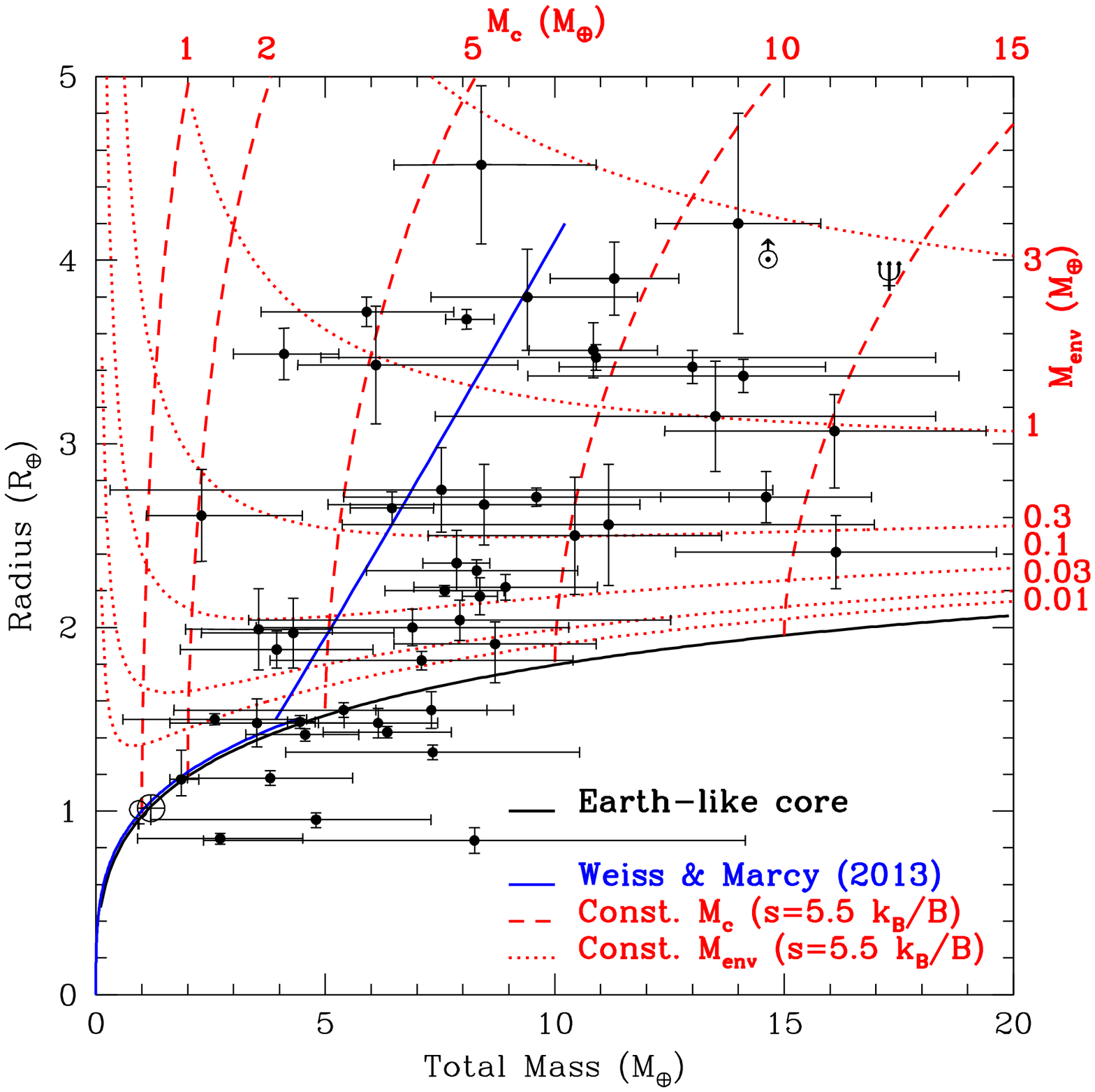}}
\subfigure{\includegraphics[width=0.49\textwidth]{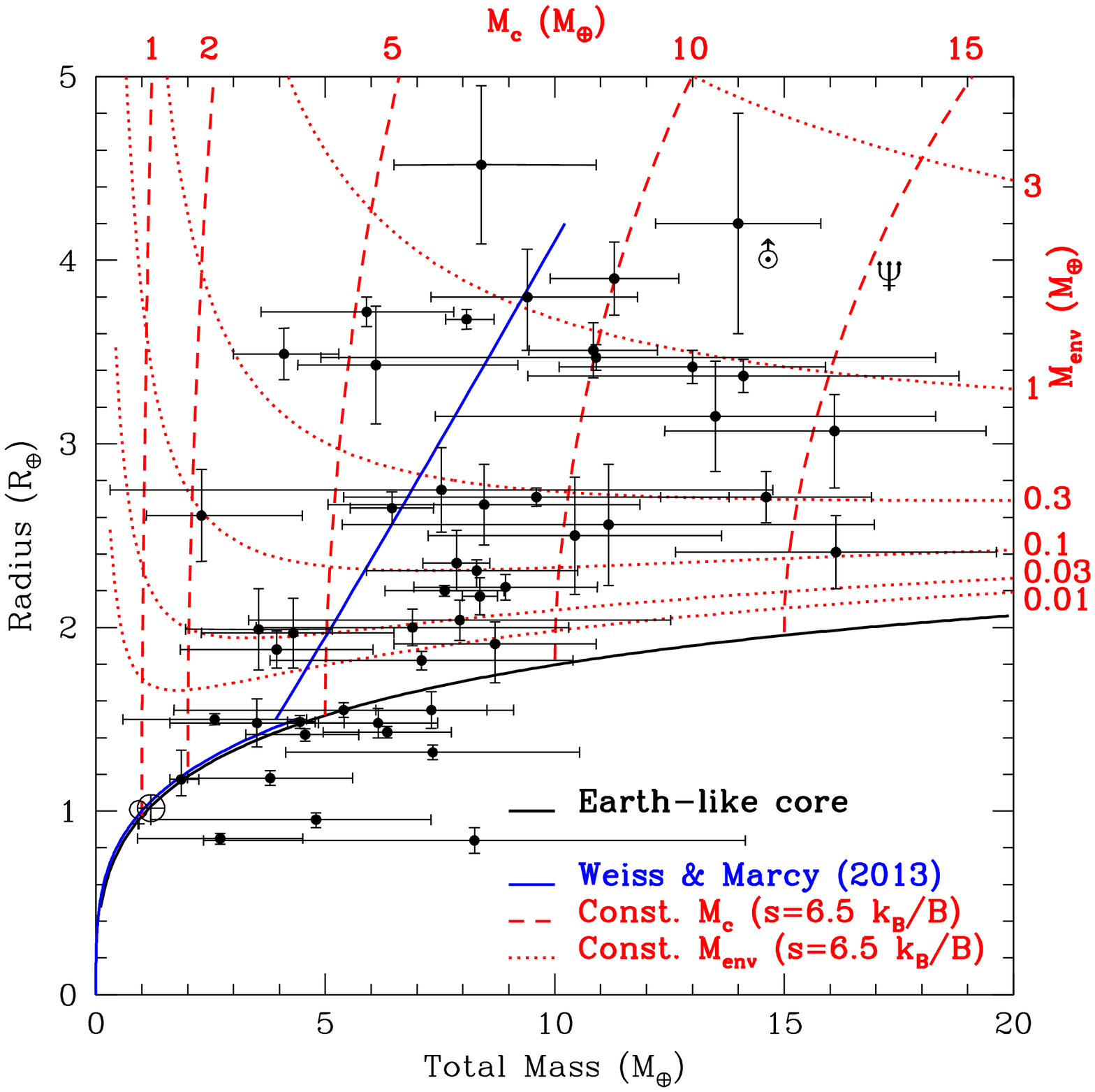}}
\end{center}
\caption{Known extrasolar planets plotted against the observational mass-radius fits of \citet{2014ApJ...783L...6W} (blue) and a grid of constant core mass and envelope mass curves (red) for planets with Earth-like (iron-silicate) cores. Curves with entropies of 5.5 and 6.5 $k_B/B$ are plotted.}
\label{planets_fits}
\end{figure*}

\begin{figure*}[htp]			
\begin{center}
\subfigure{\includegraphics[width=0.49\textwidth]{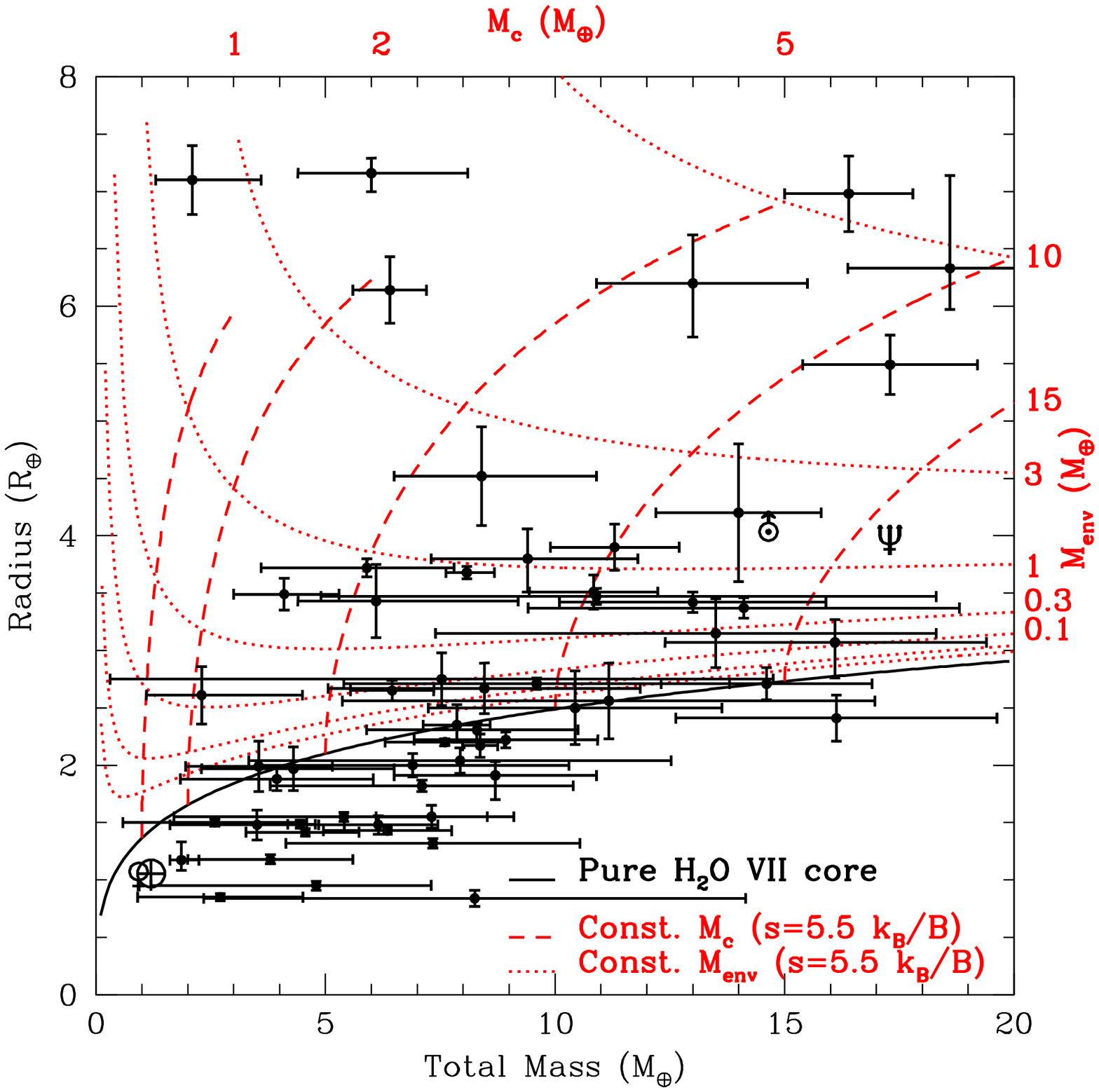}}
\subfigure{\includegraphics[width=0.49\textwidth]{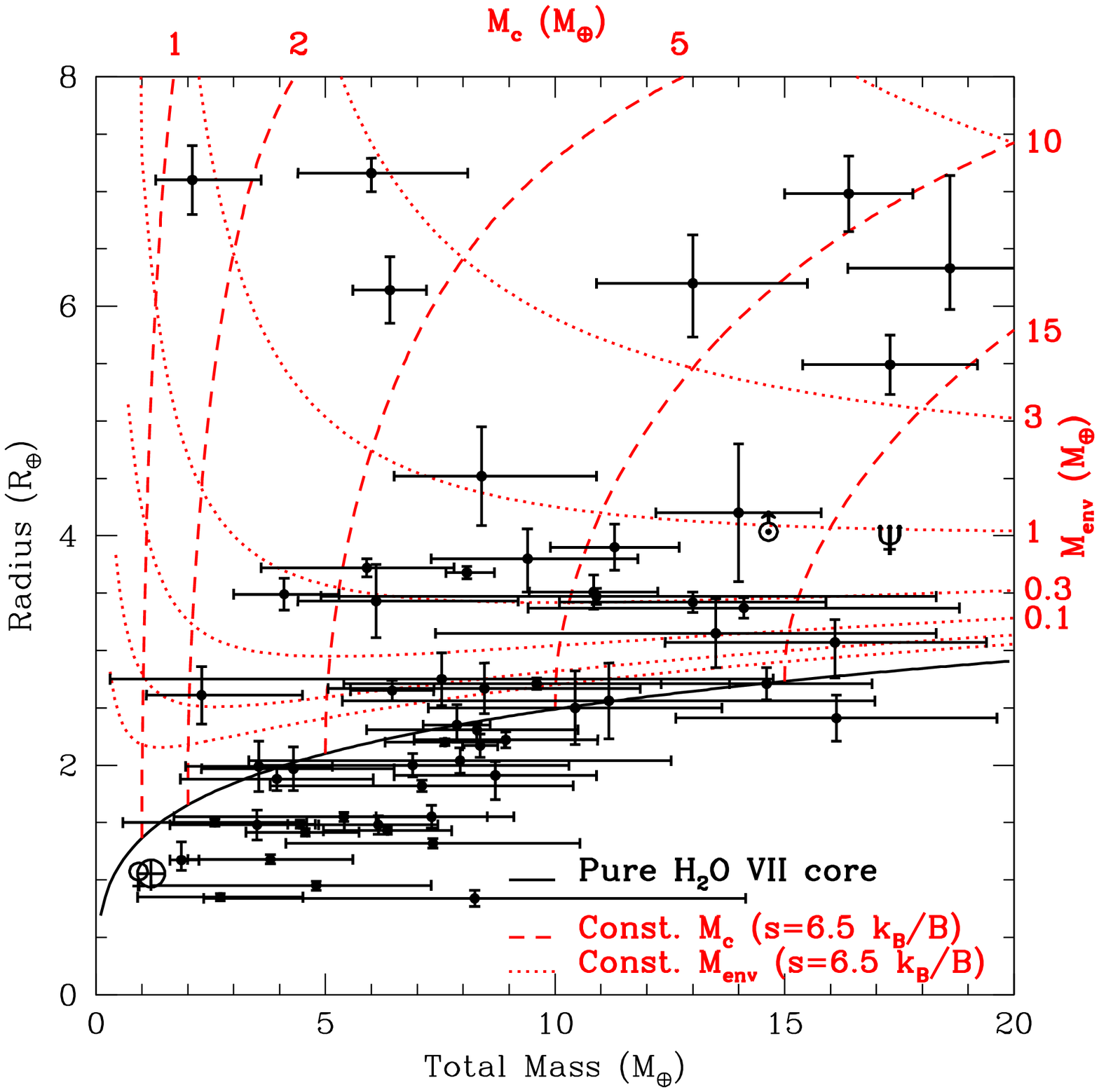}}
\end{center}
\caption{Same as Figure \ref{planets_fits}, but for pure ice cores. The functional fit has been omitted.}
\label{planets_h2o_fits}
\end{figure*}

\begin{table*}[tbhp]
\caption{Masses and Radii of Observed Exoplanets and Theoretical Decomposition into Core and Convective Envelope Components ($s = 6.0 k_B/B$)}
\begin{center}
\begin{tabular}{l|l|l|l|l|l|l|l}
\hline
Planet & Radius (R$_\earth$) & Mass (M$_\earth$) & $M_c$ (M$_\earth$) & $M_{env}$ (M$_\earth$) & $M_c$ (M$_\earth$) & $M_{env}$ (M$_\earth$) & References \\
       &                     &                   & (Fe/Rock Core)     & (Fe/Rock Core)         & (H$_2$O Core)      & (H$_2$O Core) & \\
\hline
55 Cancri e & $2.17\pm 0.10$            & $8.37\pm 0.38$         &  8.284 & 0.086 &  8.37   & 0.00   & 1,2        \\
CoRoT-7b    & $1.55\pm 0.10$            & $7.31\pm 1.21$         &  7.31  & 0.00  &  7.31   & 0.00   & 3          \\
GJ 1214b    & $2.65\pm 0.09$            & $6.45\pm 0.91$         &  6.15  & 0.30  &  6.396  & 0.054  & 4          \\
GJ 3470b    & $4.20\pm 0.60$            & $14.0\pm 1.8$          & 11.6   & 2.4   & 12.5    & 1.5    & 5          \\
HAT-P-26b   & $6.33_{-0.36}^{+0.81}$    & $18.60\pm 2.22$        &  9.9   & 8.7   & 10.9    & 7.7    & 6          \\
HD 97658b   & $2.35_{-0.15}^{+0.18}$    & $7.86\pm 0.73$         &  7.50  & 0.36  &  7.86   & 0.00   & 7          \\
Kepler-10b  & $1.416_{-0.036}^{+0.033}$ & $4.56_{-1.29}^{+1.17}$ &  4.56  & 0.00  &  4.56   & 0.00   & 8          \\
Kepler-11b  & $1.97\pm 0.19$            & $4.3_{-2.0}^{+2.2}$    &  4.247 & 0.053 &  4.3    & 0.00   & 9          \\
Kepler-11c  & $3.15\pm 0.30$            & $13.5_{-6.1}^{+4.8}$   & 12.66  & 0.84  & 13.32   & 0.18   & 9          \\
Kepler-11d  & $3.43\pm 0.32$            & $6.1_{-1.7}^{+3.1}$    &  5.35  & 0.75  &  5.67   & 0.43   & 9          \\
Kepler-11e  & $4.52\pm 0.43$            & $8.4_{-1.9}^{+2.5}$    &  6.3   & 2.1   &  6.8    & 1.6    & 9          \\
Kepler-11f  & $2.61\pm 0.25$            & $2.3_{-1.2}^{+2.2}$    &  2.14  & 0.16  &  2.226  & 0.074  & 9          \\
Kepler-18b  & $2.00\pm 0.10$            & $6.9\pm 3.4$           &  6.856 & 0.044 &  6.9    & 0.00   & 10         \\
Kepler-18c  & $5.49\pm 0.26$            & $17.3\pm 1.9$          & 11.5   & 5.8   & 12.5    & 4.8    & 10         \\
Kepler-18d  & $6.98\pm 0.33$            & $16.4\pm 1.4$          &        &       &  7.3    & 9.1    & 10         \\
Kepler-20b  & $1.91_{-0.21}^{+0.12}$    & $8.7\pm 2.2$           &  8.688 & 0.012 &  8.7    & 0.00   & 11         \\
Kepler-20c  & $3.07_{-0.31}^{+0.20}$    & $16.1_{-3.7}^{+3.3}$   & 15.30  & 0.80  & 16.002  & 0.098  & 11         \\
Kepler-20d  & $2.75\pm 0.23$            & $7.53\pm 7.22$         &  7.16  & 0.37  &  7.461  & 0.069  & 12         \\
Kepler-25b  & $2.71\pm 0.05$            & $9.6\pm 4.2$           &  9.23  & 0.37  &  9.572  & 0.028  & 12         \\
Kepler-30b  & $3.90\pm 0.20$            & $11.3\pm 1.4$          &  9.6   & 1.7   & 10.31   & 0.99   & 13         \\
Kepler-36b  & $1.486\pm 0.035$          & $4.45_{-0.27}^{+0.33}$ &  4.45  & 0.00  &  4.45   & 0.00   & 14         \\
Kepler-36c  & $3.679\pm 0.054$          & $8.08_{-0.46}^{+0.60}$ &  6.96  & 1.12  &  7.41   & 0.67   & 14         \\
Kepler-48b  & $1.88\pm 0.10$            & $3.94\pm 2.10$         &  3.902 & 0.038 &  3.94   & 0.00   & 12         \\
Kepler-48c  & $2.71\pm 0.14$            & $14.61\pm 2.30$        & 14.21  & 0.40  & 14.61   & 0.00   & 12         \\
Kepler-48d  & $2.04\pm 0.11$            & $7.93\pm 4.60$         &  7.883 & 0.047 &  7.93   & 0.00   & 12         \\
Kepler-50b  & $2.20\pm 0.03$            & $7.6\pm 1.3$           &  7.5   & 0.10  &  7.6    & 0.00   & 15         \\
Kepler-51b  & $7.10\pm 0.30$            & $2.1_{-0.80}^{+1.50}$  &        &       &  0.84   & 1.26   & 16         \\
Kepler-57c  & $1.55\pm 0.04$            & $5.4\pm 3.7$           &  5.4   & 0.00  &  5.4    & 0.00   & 15         \\
Kepler-68b  & $2.31_{-0.09}^{+0.06}$    & $8.3_{-2.4}^{+2.2}$    &  8.16  & 0.14  &  8.3    & 0.00   & 17         \\
Kepler-68c  & $0.953_{-0.042}^{+0.037}$ & $4.8_{-3.6}^{+2.5}$    &  4.8   & 0.00  &  4.8    & 0.00   & 17         \\
Kepler-78b  & $1.173_{-0.089}^{+0.159}$ & $1.86_{-0.25}^{0.38}$  &  1.86  & 0.00  &  1.86   & 0.00   & 18         \\
Kepler-79b  & $3.47\pm 0.07$            & $10.9_{-6.0}^{7.4}$    &  9.79  & 1.11  & 10.4    & 0.50   & 19         \\
Kepler-79c  & $3.72\pm 0.08$            & $5.9_{-2.3}^{1.9}$     &  4.97  & 0.93  &  5.28   & 0.62   & 19         \\
Kepler-79d  & $7.16_{-0.16}^{+0.13}$    & $6.0_{-1.6}^{2.1}$     &        &       &  2.0    & 4.0    & 19         \\
Kepler-79e  & $3.49\pm 0.14$            & $4.1_{-1.1}^{1.2}$     &  3.51  & 0.59  &  3.71   & 0.39   & 19         \\
Kepler-87c  & $6.14\pm 0.29$            & $6.4\pm 0.8$           &        &       &  3.3    & 3.1    & 20         \\
Kepler-89c  & $3.80_{-0.29}^{+0.26}$    & $9.4_{-2.1}^{-2.4}$    &  8.0   & 1.4   &  8.58   & 0.82   & 21         \\
Kepler-89e  & $6.20_{-0.47}^{+0.42}$    & $13.0_{-2.1}^{-2.5}$   &        &       &  7.3    & 5.7    & 21         \\
Kepler-93b  & $1.50\pm 0.03$            & $2.59\pm 2.00$         &  2.585 & 0.005 &  2.59   & 0.00   & 12         \\
Kepler-94b  & $3.51\pm 0.15$            & $10.84\pm 1.40$        &  9.69  & 1.15  & 10.30   & 0.54   & 12         \\
Kepler-95b  & $3.42\pm 0.09$            & $13.0\pm 2.9$          & 11.85  & 1.15  & 12.57   & 0.43   & 12         \\
Kepler-96b  & $2.67\pm 2.22$            & $8.46\pm 3.40$         &  8.12  & 0.34  &  8.428  & 0.032  & 12         \\
Kepler-97b  & $1.48\pm 0.13$            & $3.51\pm 1.90$         &  3.509 & 0.001 &  3.51   & 0.00   & 12         \\
Kepler-98b  & $1.99\pm 0.22$            & $3.55\pm 1.60$         &  3.491 & 0.059 &  3.5496 & 0.0004 & 12         \\
Kepler-99b  & $1.48\pm 0.08$            & $6.15\pm 1.30$         &  6.15  & 0.00  &  6.15   & 0.00   & 12         \\
Kepler-100b & $1.32\pm 0.04$            & $7.34\pm 3.20$         &  7.34  & 0.00  &  7.34   & 0.00   & 12         \\
Kepler-102b & $1.18\pm 0.04$            & $3.8\pm 1.8$           &  3.8   & 0.00  &  3.8    & 0.00   & 12         \\
Kepler-102e & $2.22\pm 0.07$            & $8.93\pm 2.00$         &  8.83  & 0.10  &  8.93   & 0.00   & 12         \\
Kepler-103b & $3.37\pm 0.09$            & $14.11\pm 4.70$        & 12.97  & 1.14  & 13.75   & 0.36   & 12         \\
Kepler-106c & $2.50\pm 0.32$            & $10.44\pm 3.20$        & 10.20  & 0.24  & 10.44   & 0.00   & 12         \\
Kepler-106e & $2.56\pm 0.33$            & $11.17\pm 5.80$        & 10.89  & 0.28  & 11.17   & 0.00   & 12         \\
Kepler-113b & $1.82\pm 0.05$            & $7.1\pm 3.3$           &  7.091 & 0.009 &  7.1    & 0.00   & 12         \\
Kepler-131b & $2.41\pm 0.20$            & $16.13\pm 3.50$        & 15.98  & 0.15  & 16.13   & 0.00   & 12         \\
Kepler-131c & $0.84\pm 0.07$            & $8.25\pm 5.90$         &  8.25  & 0.00  &  8.25   & 0.00   & 12         \\
Kepler-406b & $1.43\pm 0.03$            & $6.35\pm 1.40$         &  6.35  & 0.00  &  6.35   & 0.00   & 12         \\
Kepler-406c & $0.85\pm 0.03$            & $2.71\pm 0.80$         &  2.71  & 0.00  &  2.71   & 0.00   & 12         \\
Uranus      & 4.007                     & 14.536                 & 12.436 & 2.1   & 13.336  & 1.2    &            \\
Neptune     & 3.883                     & 17.147                 & 15.047 & 2.1   & 16.147  & 1.0    &            \\
\hline
\end{tabular}
\end{center}
\tablerefs{
(1) \citet{2012yCat..35399028G}
(2) \citet{2012ApJ...759...19E};
(3) \citet{2013Icar..226.1625M};
(4) \citet{2011ApJ...730...82C};
(5) \citet{2013A&A...559A..33C};
(6) \citet{2011ApJ...728..138H};
(7) \citet{2013ApJ...772L...2D};
(8) \citet{2011ApJ...729...27B};
(9) \citet{2011Natur.470...53L};
(10) \citet{2011ApJS..197....7C};
(11) \citet{2012ApJ...749...15G};
(12) \citet{2014ApJ...783L...6W};
(13) \citet{2012Natur.487..449S};
(14) \citet{2012Sci...337..556C};
(15) \citet{2013MNRAS.428.1077S};
(16) \citet{2014arXiv1401.2885M};
(17) \citet{2013ApJ...766...40G};
(18) \citet{2013arXiv1310.7987P};
(19) \citet{2013arXiv1310.2642J};
(20) \citet{2014A&A...561A.103O};
(21) \citet{2013ApJ...778..185M};
}
\label{planet_list}
\end{table*}

\begin{table*}[tbhp]
\caption{Masses and Radii of Observed Exoplanets and Theoretical Decomposition into Core and Convective Envelope Components ($s = 6.0 k_B/B$) with Radii Corrected for the Thickness ($\Delta R$) of the Radiative Atmosphere}
\begin{center}
\begin{tabular}{l|l|l|l|l|l|l|l|l}
\hline
Planet & Radius (R$_\earth$) & Mass (M$_\earth$) & $\Delta R$ (R$_\earth$) & $M_c$ (M$_\earth$) & $M_{env}$ (M$_\earth$) & $M_c$ (M$_\earth$) & $M_{env}$ (M$_\earth$) & References \\
       &                     &                   &                         & (Fe/Rock Core)     & (Fe/Rock Core)         & (H$_2$O Core)      & (H$_2$O Core) & \\
\hline
55 Cancri e & $2.17\pm 0.10$            & $8.37\pm 0.38$         & 0.523 &  8.37  & 0.00  &  8.37   & 0.00   & 1,2        \\
CoRoT-7b    & $1.55\pm 0.10$            & $7.31\pm 1.21$         & 0.284 &  7.31  & 0.00  &  7.31   & 0.00   & 3          \\
GJ 1214b    & $2.65\pm 0.09$            & $6.45\pm 0.91$         & 0.294 &  6.27  & 0.18  &  6.444  & 0.006  & 4          \\
GJ 3470b    & $4.20\pm 0.60$            & $14.0\pm 1.8$          & 0.370 & 12.1   & 1.9   & 12.98   & 1.02   & 5          \\
HAT-P-26b   & $6.33_{-0.36}^{+0.81}$    & $18.60\pm 2.22$        & 1.027 & 12.7   & 5.9   & 13.9    & 4.7    & 6          \\
HD 97658b   & $2.35_{-0.15}^{+0.18}$    & $7.86\pm 0.73$         & 0.246 &  7.782 & 0.078 &  7.86   & 0.00   & 7          \\
Kepler-10b  & $1.416_{-0.036}^{+0.033}$ & $4.56_{-1.29}^{+1.17}$ & 0.455 &  4.56  & 0.00  &  4.56   & 0.00   & 8          \\
Kepler-11b  & $1.97\pm 0.19$            & $4.3_{-2.0}^{+2.2}$    & 0.405 &  4.297 & 0.003 &  4.3    & 0.00   & 9          \\
Kepler-11c  & $3.15\pm 0.30$            & $13.5_{-6.1}^{+4.8}$   & 0.301 & 12.94  & 0.56  & 13.469  & 0.031  & 9          \\
Kepler-11d  & $3.43\pm 0.32$            & $6.1_{-1.7}^{+3.1}$    & 0.660 &  5.71  & 0.39  &  5.98   & 0.12   & 9          \\
Kepler-11e  & $4.52\pm 0.43$            & $8.4_{-1.9}^{+2.5}$    & 0.756 &  7.1   & 1.3   &  7.55   & 0.85   & 9          \\
Kepler-11f  & $2.61\pm 0.25$            & $2.3_{-1.2}^{+2.2}$    & 0.798 &  2.257 & 0.043 &  2.298  & 0.002  & 9          \\
Kepler-18b  & $2.00\pm 0.10$            & $6.9\pm 3.4$           & 0.356 &  6.9   & 0.00  &  6.9    & 0.00   & 10         \\
Kepler-18c  & $5.49\pm 0.26$            & $17.3\pm 1.9$          & 0.826 & 13.4   & 3.9   & 14.5    & 2.8    & 10         \\
Kepler-18d  & $6.98\pm 0.33$            & $16.4\pm 1.4$          & 1.128 &  9.5   & 6.9   & 10.4    & 6.0    & 10         \\
Kepler-20b  & $1.91_{-0.21}^{+0.12}$    & $8.7\pm 2.2$           & 0.241 &  8.7   & 0.00  &  8.7    & 0.00   & 11         \\
Kepler-20c  & $3.07_{-0.31}^{+0.20}$    & $16.1_{-3.7}^{+3.3}$   & 0.234 & 15.44  & 0.56  & 16.095  & 0.005  & 11         \\
Kepler-20d  & $2.75\pm 0.23$            & $7.53\pm 7.22$         & 0.209 &  7.25  & 0.28  &  7.506  & 0.024  & 12         \\
Kepler-25b  & $2.71\pm 0.05$            & $9.6\pm 4.2$           & 0.517 &  9.501 & 0.099 &  9.6    & 0.00   & 12         \\
Kepler-30b  & $3.90\pm 0.20$            & $11.3\pm 1.4$          & 0.383 & 10.0   & 1.3   & 10.69   & 0.61   & 13         \\
Kepler-36b  & $1.486\pm 0.035$          & $4.45_{-0.27}^{+0.33}$ & 0.254 &  4.45  & 0.00  &  4.45   & 0.00   & 14         \\
Kepler-36c  & $3.679\pm 0.054$          & $8.08_{-0.46}^{+0.60}$ & 0.815 &  7.58  & 0.50  &  7.95   & 0.13   & 14         \\
Kepler-48b  & $1.88\pm 0.10$            & $3.94\pm 2.10$         & 0.430 &  3.9398 & 0.0002 & 3.94  & 0.00   & 12         \\
Kepler-48c  & $2.71\pm 0.14$            & $14.61\pm 2.30$        & 0.260 & 14.40  & 0.21  & 14.61   & 0.00   & 12         \\
Kepler-48d  & $2.04\pm 0.11$            & $7.93\pm 4.60$         & 0.135 &  7.909 & 0.021 &  7.93   & 0.00   & 12         \\
Kepler-50b  & $2.20\pm 0.03$            & $7.6\pm 1.3$           &       &        &       &         &        & 15         \\
Kepler-51b  & $7.10\pm 0.30$            & $2.1_{-0.80}^{+1.50}$  &       &        &       &         &        & 16         \\
Kepler-57c  & $1.55\pm 0.04$            & $5.4\pm 3.7$           &       &        &       &         &        & 15         \\
Kepler-68b  & $2.31_{-0.09}^{+0.06}$    & $8.3_{-2.4}^{+2.2}$    & 0.384 &  8.277 & 0.023 &  8.3    & 0.00   & 17         \\
Kepler-68c  & $0.953_{-0.042}^{+0.037}$ & $4.8_{-3.6}^{+2.5}$    & 0.093 &  4.8   & 0.00  &  4.8    & 0.00   & 17         \\
Kepler-78b  & $1.173_{-0.089}^{+0.159}$ & $1.86_{-0.25}^{0.38}$  & 0.733 &  1.86  & 0.00  &  1.86   & 0.00   & 18         \\
Kepler-79b  & $3.47\pm 0.07$            & $10.9_{-6.0}^{7.4}$    & 0.522 & 10.28  & 0.62  & 10.78   & 0.12   & 19         \\
Kepler-79c  & $3.72\pm 0.08$            & $5.9_{-2.3}^{1.9}$     & 0.877 &  5.48  & 0.42  &  5.74   & 0.16   & 19         \\
Kepler-79d  & $7.16_{-0.16}^{+0.13}$    & $6.0_{-1.6}^{2.1}$     & 2.586 &  4.3   & 1.7   &  4.5    & 1.5    & 19         \\
Kepler-79e  & $3.49\pm 0.14$            & $4.1_{-1.1}^{1.2}$     & 0.776 &  3.80  & 0.30  &  3.97   & 0.13   & 19         \\
Kepler-87c  & $6.14\pm 0.29$            & $6.4\pm 0.8$           & 1.456 &  4.5   & 1.9   &  4.7    & 1.7    & 20         \\
Kepler-89c  & $3.80_{-0.29}^{+0.26}$    & $9.4_{-2.1}^{-2.4}$    & 0.799 &  8.76  & 0.64  &  9.21   & 0.19   & 21         \\
Kepler-89e  & $6.20_{-0.47}^{+0.42}$    & $13.0_{-2.1}^{-2.5}$   & 0.887 &  8.4   & 4.6   &  9.1    & 3.9    & 21         \\
Kepler-93b  & $1.50\pm 0.03$            & $2.59\pm 2.00$         & 0.444 &  2.59  & 0.00  &  2.59   & 0.00   & 12         \\
Kepler-94b  & $3.51\pm 0.15$            & $10.84\pm 1.40$        & 0.574 & 10.23  & 0.61  & 10.72   & 0.12   & 12         \\
Kepler-95b  & $3.42\pm 0.09$            & $13.0\pm 2.9$          & 0.438 & 12.31  & 0.69  & 12.90   & 0.10   & 12         \\
Kepler-96b  & $2.67\pm 2.22$            & $8.46\pm 3.40$         & 0.328 &  8.29  & 0.17  &  8.46   & 0.00   & 12         \\
Kepler-97b  & $1.48\pm 0.13$            & $3.51\pm 1.90$         & 0.447 &  3.51  & 0.00  &  3.51   & 0.00   & 12         \\
Kepler-98b  & $1.99\pm 0.22$            & $3.55\pm 1.60$         & 0.937 &  3.55  & 0.00  &  3.55   & 0.00   & 12         \\
Kepler-99b  & $1.48\pm 0.08$            & $6.15\pm 1.30$         & 0.146 &  6.15  & 0.00  &  6.15   & 0.00   & 12         \\
Kepler-100b & $1.32\pm 0.04$            & $7.34\pm 3.20$         & 0.147 &  7.34  & 0.00  &  7.34   & 0.00   & 12         \\
Kepler-102b & $1.18\pm 0.04$            & $3.8\pm 1.8$           & 0.145 &  3.8   & 0.00  &  3.8    & 0.00   & 12         \\
Kepler-102e & $2.22\pm 0.07$            & $8.93\pm 2.00$         & 0.149 &  8.873 & 0.057 &  8.93   & 0.00   & 12         \\
Kepler-103b & $3.37\pm 0.09$            & $14.11\pm 4.70$        & 0.354 & 13.37  & 0.74  & 14.01   & 0.10   & 12         \\
Kepler-106c & $2.50\pm 0.32$            & $10.44\pm 3.20$        & 0.241 & 10.32  & 0.12  & 10.44   & 0.00   & 12         \\
Kepler-106e & $2.56\pm 0.33$            & $11.17\pm 5.80$        & 0.156 & 10.97  & 0.20  & 11.17   & 0.00   & 12         \\
Kepler-113b & $1.82\pm 0.05$            & $7.1\pm 3.3$           & 0.175 &  7.1   & 0.00  &  7.1    & 0.00   & 12         \\
Kepler-131b & $2.41\pm 0.20$            & $16.13\pm 3.50$        & 0.139 & 16.047 & 0.083 & 16.13   & 0.00   & 12         \\
Kepler-131c & $0.84\pm 0.07$            & $8.25\pm 5.90$         & 0.026 &  8.25  & 0.00  &  8.25   & 0.00   & 12         \\
Kepler-406b & $1.43\pm 0.03$            & $6.35\pm 1.40$         & 0.221 &  6.35  & 0.00  &  6.35   & 0.00   & 12         \\
Kepler-406c & $0.85\pm 0.03$            & $2.71\pm 0.80$         & 0.146 &  2.71  & 0.00  &  2.71   & 0.00   & 12         \\
Uranus      & 4.007                     & 14.536                 & 0.033 & 12.436 & 2.1   & 13.336  & 1.2    &            \\
Neptune     & 3.883                     & 17.147                 & 0.021 & 15.047 & 2.1   & 16.147  & 1.0    &            \\
\hline
\end{tabular}
\end{center}
\tablerefs{
(1) \citet{2012yCat..35399028G}
(2) \citet{2012ApJ...759...19E};
(3) \citet{2013Icar..226.1625M};
(4) \citet{2011ApJ...730...82C};
(5) \citet{2013A&A...559A..33C};
(6) \citet{2011ApJ...728..138H};
(7) \citet{2013ApJ...772L...2D};
(8) \citet{2011ApJ...729...27B};
(9) \citet{2011Natur.470...53L};
(10) \citet{2011ApJS..197....7C};
(11) \citet{2012ApJ...749...15G};
(12) \citet{2014ApJ...783L...6W};
(13) \citet{2012Natur.487..449S};
(14) \citet{2012Sci...337..556C};
(15) \citet{2013MNRAS.428.1077S};
(16) \citet{2014arXiv1401.2885M};
(17) \citet{2013ApJ...766...40G};
(18) \citet{2013arXiv1310.7987P};
(19) \citet{2013arXiv1310.2642J};
(20) \citet{2014A&A...561A.103O};
(21) \citet{2013ApJ...778..185M};
}
\label{planet_list_rad}
\end{table*}

However, \citet{2014ApJ...783L...6W} find an RMS deviation in mass around their fit of 4.3 M$_\earth$, which can easily lead to a factor of two or more variation in total mass at a given radius. To address this, in Figures \ref{planets_fits} and \ref{planets_h2o_fits}, we demonstrate the power of the core-envelope decomposition by plotting known planets on a grid of constant-core-mass and constant-envelope-mass curves (red) for two entropies of 5.5 and 6.5 $k_B$ per baryon. In Figure \ref{planets_fits}, we employ models with Earth-like cores, and in Figure \ref{planets_h2o_fits}, we employ models with pure ice cores. With this grid, each mass-radius pair can be associated with a unique core mass and envelope mass for a given entropy and core type. The functional fit from \citet{2014ApJ...783L...6W} is also shown in blue in Figure \ref{planets_fits}.

Because of the larger radii of the ice cores, it takes significantly less envelope mass to produce the same radius for a given total mass. This allows us to extend Figure \ref{planets_h2o_fits} to larger radii to reflect this, noting the detection of several more planets with large radii and envelope masses of $\sim 1-10$ M$_\earth$ in this case.

Figures \ref{planets_fits} and \ref{planets_h2o_fits} assume a fully convective envelope, which is a good approximation when irradiation is low, but not when irradiation is high, and the radiative atmosphere is deep. Depending on the irradation and surface gravity, the radiative atmosphere typically comprises 10\%-20\% of the planetary radius, significantly greater than the depth of a convective atmosphere reaching the pressure of the radiative-convective boundary, which results in a larger radius than a fully-convective envelope would suggest. \citet{2014ApJ...783L...6W} find a good fit to an approximation setting the depth of the radiative atmosphere to nine times the scale height (a radiative-convective boundary at 162 bar), and we apply this approximation to estimate the effect of including the radiative atmosphere in our model. The actual depth of the radiative-convective boundary will depend on irradiation, age, and metallicity.

We provide the quantitative core-envelope decomposition for observed exoplanets and Solar System planets in Tables \ref{planet_list} and \ref{planet_list_rad} in both the rocky-iron core and ice core cases for an envelope entropy of 6.0 $k_B$ per baryon. Tables \ref{planet_list} gives the decomposition for a fully-convective envelope, and Table \ref{planet_list_rad} gives the decomposition with the correction for the radiative atmosphere included. If the observed radius is smaller than a bare core of the observed mass for one or both core types, we still include the decomposition with an envelope mass of zero. Uranus and Neptune are included with both core types for comparison purposes.

The core-envelope decompositions readily reveal useful information about the structures of planets from mass-radius data. For example, in the fully convective case, the low-density planet Kepler-11e (the topmost point plotted in Figure \ref{planets_fits}), can be fit to a model with a large H$_2$-He fraction based on an Earth-like core mass of 6.3 M$_\earth$ and an envelope mass of 2.1 M$_\earth$, for an entropy for 6.0 $k_B$ per baryon, with small differences for a different entropy\footnote{Increasing the entropy of the models shifts the grid up and to the left, making it straighter, and fits a higher core mass and lower envelope mass to the same mass and radius.}, or an icy core mass of 6.8 M$_\earth$ and envelope mass of 1.6 M$_\earth$. In contrast, the higher-density planet Kepler-131b (the bottom-rightmost point) can be fit to a model with a small H$_2$-He fraction based on an Earth-like core mass of 15.98 M$_\earth$ and an envelope mass of 0.15 M$_\earth$ at an entropy of 6.0 $k_B$ per baryon. For an ice core, the bare core would suffice.

The difference in envelope mass between an Earth-like core model and an ice core model can be as small as 0.059 M$_\earth$ while still retaining some envelope (e.g. Kepler-98b), but is usually a few tenths of an Earth mass. The largest difference of 1.1 M$_\earth$ occurs for Neptune.

When the correction for the radiative atmosphere is applied, the H$_2$-He mass fractions required to fit the measured masses and radii decrease significantly. A number of planets can be modeled as a solid core with only a radiative atmosphere (which in all cases has a mass fractions of $<10^{-4}$) and no convective envelope. At the other extreme, when modeled with an ice core, the low-density planet Kepler-18d has a hydrogen fraction of 55\% assuming a fully convective envelope, but 35\% given the correction for the radiative atmosphere, the largest H$_2$-He mass fraction of any planet we study in the latter case. The results given by \citet{2012ApJ...761...59L} for the Kepler-11 system are consistent with these corrected results for an Earth-like core, except for Kepler-11f, for which we predict a smaller envelope mass fraction.

We note a small number of planets with masses of 2$-$20 M$_\earth$ and very large radii (for these masses) of 5$-$7 R$_\earth$, such as Kepler-18d, as shown in Figure \ref{planets_h2o_fits}. While additional planet detections and further investigation with gas giant models is needed to investigate these objects in detail, we see from these objects that large envelope fractions can occur even for low-mass planets. Even including the radiative atmosphere correction, we find H$_2$-He mass fractions of 22\%-35\% for this population. (Without this correction, the H$_2$-He mass fractions are approximately twice as large.)

This core-envelope decomposition for observed planets, especially when the large-radius planets are included, implies that envelope mass can vary from zero to tens of percent of the total mass for the entire range of masses we study, except that we find a lower limit of envelope mass of $\sim$0.1 M$_\earth$ in the case of Earth-like cores with masses of $\sim 8-20$ M$_\earth$. More specifically, where a non-zero envelope mass is predicted, it can vary by two orders of magnitude for a similar core mass. While we do see the trend of decreasing density for masses $>7.6$ M$_\earth$ observed by \citet{2014ApJ...783L...6W}, with a broad scatter of the observed planets centered around the functional fit, the spread in core mass and envelope mass is so wide that we see only a slight justification for any given functional fit.

We consider the possibility of multiple populations of planets in envelope mass space$-$one with $M_{env} \gtrsim$3 M$_\earth$, one with $M_{env} \sim 1$ M$_\earth$, and one with $M_{env} \lesssim 0.3$ M$_\earth$, but the statistics are not sufficient to tell whether these populations are distinct. In any case, it is clear that a significant amount of H$_2$-He ($\sim$1 M$_\earth$) is needed to produce the large radii observed for many low-mass planets. We also note that many of the newly-discovered planets fall in the 5$-$10 M$_\earth$ core mass range, but, again, it is not clear if this genuinely reflects the true underlying distribution function, or is due to statistics and selection biases.


\section{Uncertainty Range}
\label{error}

\begin{table*}[tbhp]
\caption{Computed Envelope Masses for Selected Planets From Different Models}
\begin{center}
\begin{tabular}{l|l|l|l|l|l|l}
\hline
           & Kepler-98b     & Kepler-11f          & Kepler-25b     & Kepler-20c             & Kepler-11e          & HAT-P-26b              \\
\hline
Mass       & $3.55\pm 1.60$ & $2.3_{-1.2}^{+2.2}$ & $9.6\pm 4.2$   & $16.1_{-3.7}^{+3.3}$   & $8.4_{-1.9}^{+2.5}$ & $18.60\pm 2.22$        \\
Radius     & $1.99\pm 0.22$ & $2.61\pm 0.25$      & $2.71\pm 0.05$ & $3.07_{-0.31}^{+0.20}$ & $4.52\pm 0.43$      & $6.33_{-0.36}^{+0.81}$ \\
$\Delta R$ & 0.937          & 0.798               & 0.517          & 0.234                  & 0.756               & 1.027                  \\
\hline
Purely Convective Models \\
\hline
Rock-Iron Core, s=5.5$^1$ & 0.086 & 0.230 & 0.47 & 0.95 & 2.6 &     \\
Rock-Iron Core, s=6.5 & 0.035  & 0.097 & 0.28  & 0.64  & 1.6  & 7.2 \\
Ice Core, s=5.5       & 0.0008 & 0.125 & 0.049 & 0.123 & 2.1  & 9.2 \\
Ice Core, s=6.5       & 0.0003 & 0.038 & 0.014 & 0.044 & 1.1  & 6.2 \\
\hline
Models With Radiative Atmsophere \\
\hline
Rock-Iron Core, s=5.5 & 0.00   & 0.061 & 0.127 & 0.66  & 1.6  & 6.7 \\
Rock-Iron Core, s=6.5 & 0.00   & 0.027 & 0.069 & 0.45  & 1.0  & 5.0 \\
Ice Core, s=5.5       & 0.00   & 0.003 & 0.00  & 0.006 & 1.10 & 5.5 \\
Ice Core, s=6.5       & 0.00   & 0.001 & 0.00  & 0.003 & 0.62 & 3.8 \\
\hline
\end{tabular}
\end{center}
$^1$Entropies are given in units of $k_B$ per baryon.
\label{error_table}
\end{table*}

All of the model parameters we consider$-$envelope entropy, core composition, and the depth of the radiative atmosphere$-$have effects on our computed core-envelope decompositions. Tables \ref{planets_fits} and \ref{planets_h2o_fits} explore the limiting cases for core composition, that is, rock-iron cores and pure ice cores, as well as the limiting cases for the correction for the radiative atmosphere, that is, no correction versus a suggested upper bound of the depth of nine scale heights, as fit by \citet{2013arXiv1311.0329L}. To provide a picture of the uncertainties involved in our modeling of exoplanets, we compute in Table \ref{error_table} the envelope masses for selected exoplanets that populate a large part of the mass-radius space, listed in order of increasing radius. In this table, we compute envelope masses for the above parameters as well as envelope entropies of 5.5 and 6.5 $k_B$ per baryon, which bracket the expected range of entropies predicted by \citet{2013arXiv1311.0329L}. In Tables \ref{planets_fits} and \ref{planets_h2o_fits}, we compute models with only $s=6.0\,k_B$ per baryon.

There is not a simple empirical rule for the trends in our results, but several important features can be seen. Most notably, we may consider very roughly two categories of objects: those with relatively small radii that can only be fit with low-mass envelopes (e.g., Kepler-98b) and those with relatively large radii that can only be fit with high-mass envelopes (e.g., HAT-P-26b). These categories are arbitrary, since the population is essentially a continuum with objects like Kepler-20c stradling the boundary, but they serve to illustrate the limits we can place on individual objects and the effects of varying the parameters. The behavior of each parameter is significantly different in the two categories.

The correction for the presence of the radiative atmosphere reduces the convective envelope masses needed to fit the mass and radius data. For planets that have especially low-mass envelopes without the correction, this can go to zero, that is, only the (much lower mass) radiative atmosphere is needed. This can be seen with 55 Cancri e (envelope mass 0.086 M$_\earth$ for $s=6.0\,k_B$ per baryon and a rock-iron core) and Kepler-98b (envelope mass 0.086 M$_\earth$ for $s=5.5\,k_B$ per baryon and a rock-iron core). However, other planets with similarly small envelopes still require a small, but non-zero envelope mass to fit the observational data, which may be more than an order of magnitude smaller than without the correction. For example, for $s=6.0\,k_B$ per baryon and a rock-iron core, Kepler-11b requires an envelope mass of 0.053 M$_\earth$ without the correction and 0.003 M$_\earth$ with it. In other cases a range of values may be required that nevertheless all remain relatively small ($\lesssim 0.5$ M$_\earth$), such as in the case of Kepler-79e, where we see envelope masses of 0.59 and 0.39 M$_\earth$ without the correction for the two different core types (rock-iron and ice, respectively), and corresponding masses of 0.30 and 0.13 M$_\earth$ with the correction. We typically find differences in envelope mass of 0.1-0.4 M$_\earth$ due to this correction for planets in this category.

On the other hand, planets for which the envelope mass is relatively large ($\gtrsim 1$ M$_\earth$) are modeled with large envelopes regardless of the parameters of the model. For example, for $s=6.0\,k_B$ per baryon, Kepler-18c requires an envelope mass of 5.8 M$_\earth$ for a rock-iron core and 4.8 M$_\earth$ for an ice core without the correction for the radiative atmosphere. With the correction, these values are 3.9 M$_\earth$ and 2.8 M$_\earth$, respectively. Similarly, without the correction, for HAT-P-26b with $s=6.0\,k_B$ per baryon, our modeled envelope mass is 8.7 M$_\earth$ for a rock-iron core and 7.7 M$_\earth$ for an ice core. With the correction, these values are 5.9 M$_\earth$ and 4.7 M$_\earth$, respectively$-$somewhat larger differences. On the other hand, without the correction, for GJ 3470b with $s=6.0\,k_B$ per baryon, our modeled envelope mass is 2.4 M$_\earth$ for a rock-iron core and 1.5 M$_\earth$ for an ice core. With the correction, these values are 1.9 M$_\earth$ and 1.02 M$_\earth$, meaning that the effect of the correction is significantly smaller in absolute terms. In general, we find that the correction for the radiative envelope reduces the envelope mass by 30\%-50\% for planets with large envelopes. The exceptions to this are Uranus and Neptune, for which the scale height is much smaller, and the radiative atmosphere makes a negligible contribution to the radius.

We can do a similar analysis of the effect of changing from a rock-iron core to a pure ice core. Because of the larger core radius, the need for a convective envelope can disappear for the ice core. If there is no radiative atmosphere, this corresponds to the bare core. The most extreme case of this is Kepler-48c, which requires an envelope mass of 0.40 M$_\earth$ to fit a rock-iron core with $s=6.0\,k_B$ per baryon and no radiative atmosphere, but no envelope to fit an ice core with otherwise the same parameters. For planets with small envelopes, we typically find differences in envelope mass of 0.2-0.5 M$_\earth$ for the different core types. For example, for Kepler-79e with $s=6.0\,k_B$ per baryon and no radiative atmosphere, the envelope mass falls from 0.59 M$_\earth$ to 0.39 M$_\earth$ when switching from a rock-iron core to an ice core. The values for the same models for Kepler-11c are 0.84 M$_\earth$ and 0.18 M$_\earth$, respectively.

For planets with large envelopes, we typically find differences in envelope mass of $\sim 1$ M$_\earth$ between models with a rock-iron core and an ice core. For example, with $s=6.0\,k_B$ per baryon and no radiative atmosphere, the respective envelope masses are 8.7 M$_\earth$ and 7.7 M$_\earth$ for HAT-P-26b, and 5.8 M$_\earth$ and 4.8$_\earth$ for Kepler-18c. With the correction for the radiative atmosphere, we find 1.9 M$_\earth$ and 1.02 M$_\earth$ for GJ 3470b, and 4.6 $M_\earth$ and 3.9 M$_\earth$ for Kepler-89e.

Table \ref{error_table} shows the effect of varying the entropy of the envelope. Varying only the entropy cannot eliminate the need for a convective envelope because this depends on the core radius and the depth of the radiative atmosphere, which do not depend on the entropy. Instead, we see the masses of small envelopes vary by a factor of 2-3 over an entropy range of 5.5 to 6.5 $k_B$ per baryon, e.g., from 0.086 M$_\earth$ to 0.035 M$_\earth$, respectively, for Kepler-98b with a rock-iron core and no radiative atmosphere, from 0.230 M$_\earth$ to 0.097 M$_\earth$ for the same parameters for Kepler-11f, and from 0.47 M$_\earth$ to 0.28 M$_\earth$ for the same parameters for Kepler-25b.

For planets with larger envelopes, the variation in envelope mass over this entropy range is 30\%-50\%. For example, for Kepler-11e with a rock-iron core and no radiative atmosphere, we find envelope masses of 2.6 M$_\earth$ for $s=5.5 k_B$ per baryon and 1.6M $_\earth$ for $6.5 k_B$ per baryon. For an ice core, we similarly find values of 2.1 M$_\earth$ and 1.1$M_\earth$, respectively. For HAT-P-26b with an ice core and the correction for the radiative atmosphere, the respective masses are 5.5 M$_\earth$ and 3.8 M$_\earth$.

We do not expect planets of Gyr ages to fall significantly outside the range of $5.5-6.5\,k_B$ per baryon. However, we find that the inferred envelope masses for a given mass and radius in our models decrease roughly linearly with increasing entropy, so that a very wide possible entropy might result in a range of envelope masses of a factor of 2, still of similar magnitude to the effects of the other uncertainties. For example, we compute envelope masses for Kepler-11e with a rock-iron core and a radiative atmosphere for an entropy range of $4.3-7.0 k_B$ per baryon. In this case, we compute envelope masses of 2.1 M$_\earth$ for an entropy of $4.3 k_B$ per baryon, 1.6 M$_\earth$ for an entropy of 5.0 $k_B$ per baryon, 1.3 M$_\earth$ for an entropy of 6.0 $k_B$ per baryon, and 0.8 M$_\earth$ for an entropy of 7.0 $k_B$ per baryon.

Table \ref{error_table} also allows us to bracket the envelope masses provided by all of our combinations of parameters, i.e., the combined effect of varying all three of the above parameters. Again, the minimum envelope mass required to fit the mass-radius data may be zero when considering all of the models, as for Kepler-98b and Kepler-25b, or small, but non-zero, e.g., 0.001 M$_\earth$ for Kepler-11f. Meanwhile, the largest envelope masses provided by any of the eight limiting cases of the variation of the three parameters (invariably $s=5.5 k_B$ per baryon, rock-iron core, and no radiative atmosphere) are 0.086 M$_\earth$ for Kepler-98b, 0.230 for Kepler-11f, and 0.47 for Kepler-25b. Even with the effects of the three parameters combined, we see a range of relatively small envelope masses, e.g. 0.00-0.47 M$_\earth$, for objects with smaller radii.

For planets with large envelopes, we see a range of envelope masses, but they are invariably relatively large. The smallest envelope mass produced by any of our models for Kepler-11e is 0.62 M$_\earth$, while the largest is 2.6 M$_\earth$. Alternatively, we may express this as an envelope mass of $1.6\pm 1.0$ M$_\earth$, a range that accounts for all of the uncertainties in the parameters of the model for Kepler-11e. Similarly the smallest envelope mass produced for HAT-P-26b is 3.8 M$_\earth$, while the largest is 9.2 M$_\earth$, so that the envelope mass of $6.0\pm 3.2$ M$_\earth$. In both cases, the uncertainty in the results is roughly similar: 63\% for Kepler-11e and 53\% for HAT-P-26b.

In summary, and importantly, even with this wide range of results for individual objects, we determine a wide range of required envelope masses (one to two orders of magnitude) for exoplanets in this important radius and total mass range.


\section{Internal Structures}
\label{internal}

\begin{figure}[htp]
\includegraphics[width=\columnwidth]{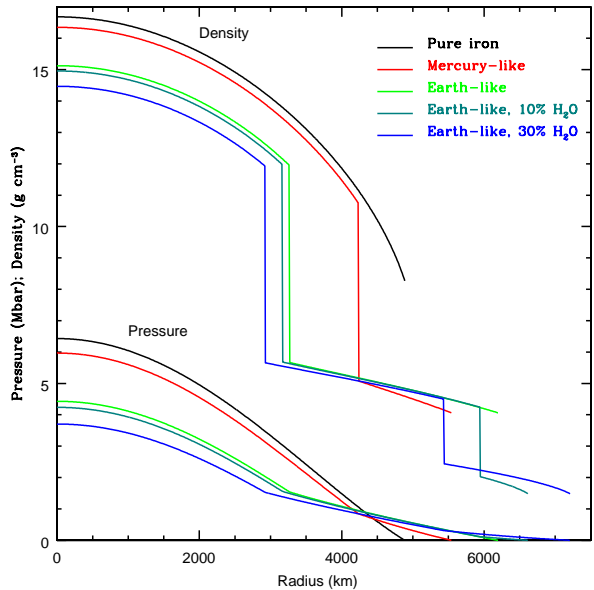}
\caption{Density and pressure profiles of various iron core/MgSiO\(_3\) mantle planets, including two with deep water layers, all 1 M$_\earth$. ``Earth-like'' is defined as a 13:27 ratio of iron to MgSiO$_3$ mass, and ``Mercury-like'' is defined as 7:3 ratio of iron to MgSiO$_3$ mass, corresponding to the compositions of Earth and Mercury when water is not included. Planets with a greater core mass fraction have a smaller radius for a given mass due to the high density of iron relative to that of MgSiO\(_3\).}
\label{density_pressure_profiles}
\end{figure}

\begin{figure}[htp]
\includegraphics[width=\columnwidth]{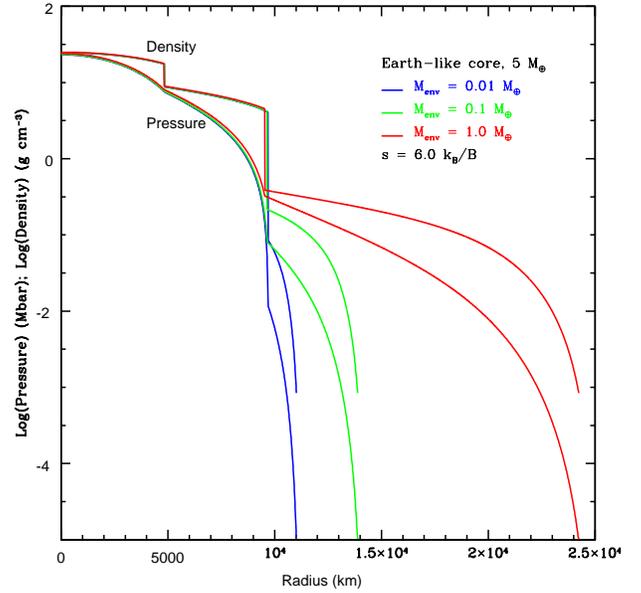}
\caption{Density and pressure profiles of planets with a 5-M$_\earth$ Earth-like core plus H$_2$-He envelopes equal to 0.01, 0.1, and 1.0 M$_\earth$. In each case, the envelope has a constant entropy of 6.0 $k_B/B$. Increasing the envelope mass compresses the core slightly, making the core radius smaller.}
\label{density_pressure_h2}
\end{figure}

In this section, we provide several figures relating to the internal structures of the planets we model. Figure \ref{density_pressure_profiles} shows pressure and density profiles of several representative solid planet models, each with a mass of 1 M$_\earth$. We include pure iron, Mercury-like, and Earth-like compositions, along with ``water-worlds'' that are otherwise Earth-like, but have 10\% and 30\% water (Ice VII). Similarly, Figure \ref{density_pressure_h2} shows pressure and density profiles of representative models with H$_2$-He envelopes. These models each have a core mass of 5 M$_\earth$ and entropy of 6.0 $k_B$ per baryon, and envelope masses of 0.01, 0.1, and 1.0 M$_\earth$ are plotted. Increasing the envelope mass slightly compresses the core, decreasing the core radius.

\begin{figure}[htp]
\includegraphics[width=\columnwidth]{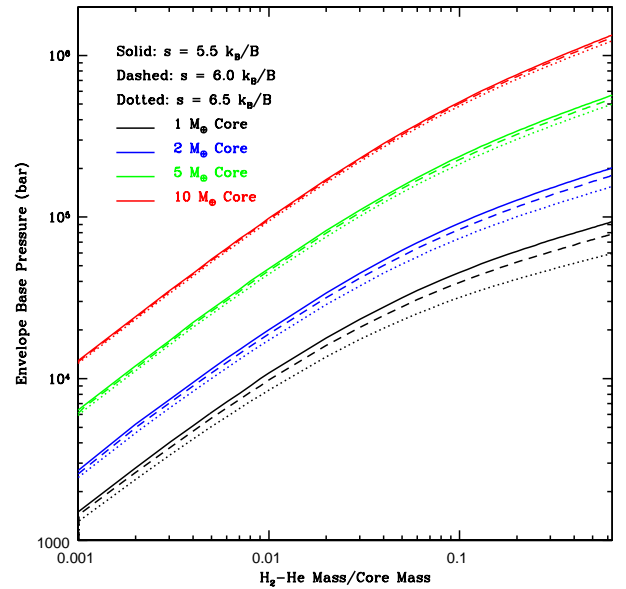}
\caption{Pressure at the base of the envelope versus envelope mass fraction for planets with constant core masses of 1, 2, 5, and 10 M$_\earth$. Curves with entropies of 5.5, 6.0, and 6.5 $k_B/B$ are plotted.}
\label{base_pressure_varh2}
\end{figure}

Figure \ref{base_pressure_varh2} shows the base pressure of the envelope versus envelope mass {\it fraction} ($f_{env}$) for models with various constant core masses. For shallow envelopes for which the gravity does not vary significantly, the expected base pressure is
\begin{equation}
P_b \approx \frac{M_{env}g}{4\pi R^2} \approx \frac{GM_cM_{env}}{4\pi R^4} \approx \frac{GM_c^2f_{env}}{4\pi R^4},
\end{equation}
where $g$ is the surface gravity. This relation holds well for small envelope mass fractions, but the pressure is lower than the relation would imply for envelope mass fractions $\gtrsim 3\%$ due to the variation in gravity over the height of the envelope.

\begin{figure}[htp]
\includegraphics[width=\columnwidth]{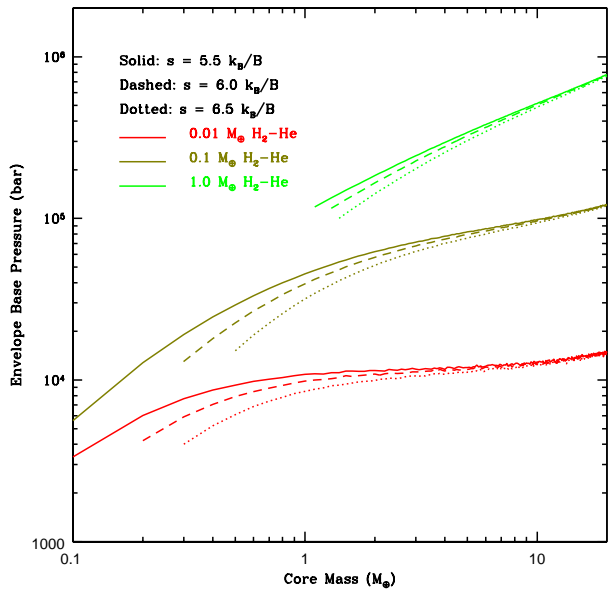}
\caption{Pressure at the base of the envelope versus core mass for planets with constant envelope masses of 0.01, 0.1, and 1.0 M$_\earth$. Curves with entropies of 5.5, 6.0, and 6.5 $k_B/B$ are plotted.}
\label{base_pressure_varcore}
\end{figure}

For low envelope fractions, the plot also shows a relation remarkably close to $P_b\propto M_c$. Figure \ref{base_pressure_varcore} shows why this is so, plotting envelope base pressure versus {\it core} mass for constant envelope masses of 0.01, 0.1, and 1.0 M$_\earth$. The envelope base pressure in nearly constant over a wide range of core masses for shallow envelopes. Evidently, based on the relation above, $M_c/R^4$ is nearly constant for these models. This is supported by our power law fit for Earth-like solid planets of $R\propto M^{0.266-0.274}$ (see Section \ref{solid}). On the other hand, if the core mass is low enough that the H$_2$-He comprises $\gtrsim$10\% of the total mass, then the base pressure increases roughly linearly with core mass. Also, the entropy begins to have a significant effect on the base pressure in models with small cores, likely due to the increasing dependence of the radius on entropy as the core mass decreases (see Figure \ref{mass_radius_varcore}).

\begin{figure}[htp]
\includegraphics[width=\columnwidth]{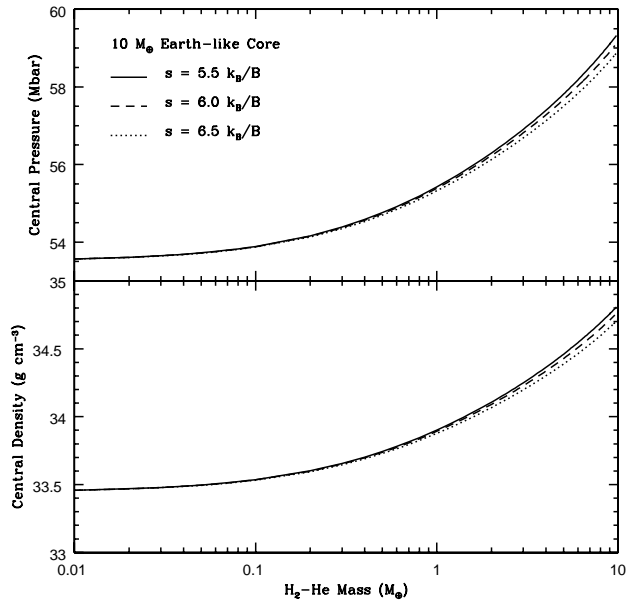}
\caption{Central pressure (top) and density (bottom) of planets with a 10-M$_\earth$ Earth-like core and a variable envelope mass. Entropies of 5.5, 6.0, and 6.5 $k_B/B$ are plotted. Despite a large change in total mass, the central pressure and density change very little.}
\label{central_pressure_density_h2}
\end{figure}

\begin{figure}[htp]
\includegraphics[width=\columnwidth]{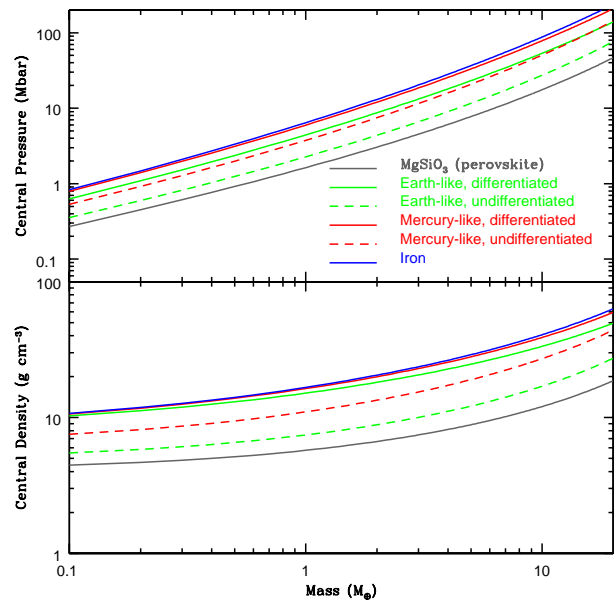}
\caption{Central pressure (top) and density (bottom) as a function of mass for constant-composition iron core/MgSiO\(_3\) (perovskite) mantle planets in the 0.1$-$20 M\(_\earth\) range. The iron mass fractions shown are 0\%, 32.5\% (Earth-like), 70\% (Mercury-like) and 100\%. Solid lines are differentiated, dashed lines are undifferentiated.}
\label{central_pressure_density}
\end{figure}

We also plot curves of central pressure and density versus envelope mass for models with a constant core mass of 10 M$_\earth$ and entropies of 5.5, 6.0, and 6.5 $k_B$ per baryon. Despite a large change in total mass, the central pressure and density change very little$-$about 5\% in density and 10\% in pressure for a doubling of total mass.

Central pressure and density versus (total) mass curves for representative solid planets, both differentiated and undifferentiated, are compared in Figure \ref{central_pressure_density}. The pure MgSiO\(_3\) (perovskite) planet profile is included in order to demonstrate the discontinuity that arises in central density when a material of higher density is added. The planets with iron cores all have very similar central densities, while undifferentiated planets with varying iron fractions have lower central densities. This pattern repeats to a lesser degree for central pressures.


\section{Solid Exoplanets}
\label{solid}

For completeness, we now provide mass-radius curves and tables for a range of solid exoplanet models with no significant envelopes. We construct true ``terrestrial'' planets with iron and MgSiO$_3$ components as well as three-component models with iron, MgSiO$_3$, and water, in the form of Ice VII, corresponding to ``water worlds.'' Figure \ref{exoplanets} shows mass-radius plots for some of our representative solid planet models compared with masses and radii of observed exoplanets.

While we compute each curve from 0.1 to 20 M$_\earth$, our core-envelope decomposition results suggest that planets larger than $\sim$8 M$_\earth$ will likely have gaseous envelopes. Therefore our models of solid planets likely should only be applied to planets $\lesssim$8 M$_\earth$ that do not have extended envelopes. However, gaseous envelopes of close-orbiting planets could later be stripped away by evaporation by XUV irradiation, particularly for planets with initial masses $\lesssim$0.3 M$_J$ orbiting young Solar-type stars \citep{2007Icar..187..358H}.

\begin{figure}[htp]
\includegraphics[width=\columnwidth]{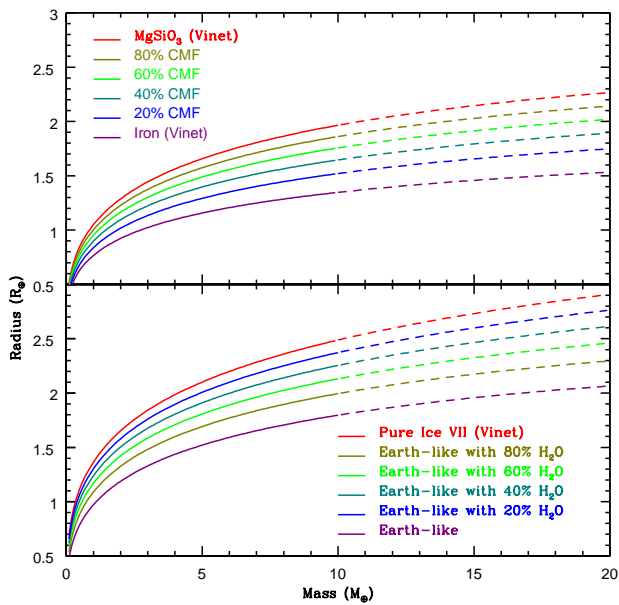}
\caption{Mass-radius curves for solid planets with various core mass fractions (CMF) ranging from 0\% to 100\%. Top panel: terrestrial models with iron cores and MgSiO$_3$ mantles. Bottom panel: ``water worlds'' with ``cores'' having an Earth-like structure plus a deep water layer (in the form of Ice VII).}
\label{mass_radius_cmf}
\end{figure}

In Figure \ref{mass_radius_cmf}, we plot terrestrial models with different iron core mass fractions (CMFs), ranging from pure iron to pure MgSiO$_3$ (perovskite), showing a systematic trend of decreasing radius with increasing CMF. We also plot models of ``water worlds'' with an Earth-like core (Earth-like in having the same Fe/Mg ratio as Earth), ranging from a purely Earth-like composition to pure water. Because the water worlds have some iron content, they overlap with terrestrial models with lower iron content, even given the lower density of water, pointing to a degeneracy with composition in the mass-radius plot. These results show good agreement with those of \citet{2007ApJ...665.1413V}, \citet{2007Icar..191..337S}, \citet{2007ApJ...669.1279S}, and \citet{2007ApJ...659.1661F}.

A power-law fit of the form \(R \propto M^x\) provides a simple description of the behavior of the mass radius curves. \citet{2006Icar..181..545V} performed such a fit over the span of their ``super-Earths'' (1$-$10 Earth masses, 33\% CMF) and ``super-Mercurys'' (1$-$10 Mercury masses, 70\% CMF), and we compute the values of \(x\) in our fits for comparison. For ``super-Earths,'' we find a power-law coefficient of 0.266$-$0.274, while \citet{2006Icar..181..545V} reported a range of 0.267$-$0.272. For ``super-Mercurys,'', we find a power-law coefficient of 0.309$-$0.312, comparable to the ``\(\sim\)0.3'' reported by \citet{2006Icar..181..545V}. The lower masses of the ``super-Mercurys'' result in less compression in their interiors, and a power-law fit closer to the \(R\propto M^{\frac{1}{3}}\) law for uncompressed planets than for super-Earths.

The super-Earths with currently observed masses and radii consistent with a purely solid composition have masses of $\sim 2-8$ M$_\earth$. While these numbers are loose, we can use them in conjunction with our models to study the relation between radius and composition. For example, in this $\sim 2-8$ M$_\earth$ range, the range of radii for pure iron models (the minimum radius for a given mass) is $\sim 0.9-1.3$ R$_\earth$, in contrast with the range for pure silicates (the maximum radius for terrestrial compositions), which have radii of $\sim 1.25-2.0$ R$_\earth$, and water-rich models, which are larger still. While there are significant degeneracies with composition among solid planets, this reiterates the usefulness of radius as a proxy for distinguishing solid exoplanets from those with gaseous envelopes, since known planets larger than 2 R$_\earth$ cannot be purely rock/iron and are likely to have such gaseous envelopes \citep{2013arXiv1311.0329L,2014ApJ...783L...6W,2014ApJS..210...20M}.\footnote{However, further work may be needed to distinguish planets with an H$_2$-He envelope from the potential population of water-rich planets.}

Properties of differentiated planet models with Earth-like and Mercury-like compositions are given in Tables \ref{0.325Fe_planet_table} and \ref{0.7Fe_planet_table}, respectively. For comparison, we present properties of pure iron and pure silicate planets in Tables \ref{pv_planet_table} and \ref{Fe_planet_table}, respectively.


\section{Conclusions}
\label{conclusions}

We have investigated a range of exoplanet models for various core masses, gaseous envelope masses, and envelope entropies, and compared them with mass and radius observations. Some of our representative modeled planet properties are tabulated in Tables \ref{0.1menv_6.0_table}-\ref{Fe_planet_table} for a variety of planet compositions. We have explored models with both ``Earth-like'' rock-iron cores and ice cores to account for the possibility of the formation of planets with gaseous envelopes in both the warm and cold regions of their ``solar'' systems, and we have investigated the correction for the presence of a radiative atmosphere. We also considered varying silicate and water fractions for solid planets.

We have decomposed observed exoplanets into core mass and envelope mass components for both rock-iron cores and ice cores. Based on measured masses and radii, we find that the envelope mass, $M_{env}$, may vary over a wide range of values from zero to tens of percent of the total mass over a wide range of total masses, except that for the higher-mass ``sub-Neptune'' (or ``mini-Neptune'') planets, a nonzero envelope mass is always required to fit a rock-iron core. Thus, planetary formation and evaporation models need to account for the very wide range of core masses and envelope masses derived.

In general, an ice core model requires a smaller envelope mass (and larger core mass) to fit the same mass-radius pair as an Earth-like core model, because of the larger radius of the core. Entropy also has a systematic effect on the core-envelope decomposition: an envelope with higher entropy is hotter and has a lower density, so that a smaller envelope mass is needed to fit the same mass-radius pair. The correction for the presence of a radiative atmosphere is also significant, reducing the envelope mass needed to fit the data by 30\%-50\% for large envelopes, and sometimes eliminating the need for a convective envelope entirely in the case of small atmosphere masses. Therefore, given these uncertainties, we have derived a range of possible core and envelope masses for known Neptune- and sub-Neptune-sized exoplanets.

While a few planets have large envelope mass fractions of $\sim 22\%-35\%$ (when corrected for the radiative atmosphere), as shown in Figure \ref{planets_h2o_fits} and Table \ref{planet_list}, most of the ``sub-Neptune'' planets that have been observed are dominated by their core masses (Figures \ref{planets_fits} and \ref{planets_h2o_fits}); that is, the envelope comprises only a small fraction of the total mass. At the same time, for core masses of $M_c\gtrsim 5$ M$_\earth$, the planetary radius is very sensitive to the envelope mass and also to the entropy, so that the observed radius can serve as a proxy for the properties of the envelope, subject to the degeneracies of envelope mass, entropy, and the depth of the radiative atmosphere.

For solid planets up to 20 R$_\earth$, we find that only planets with radii $\lesssim 1.5-2.0$ R$_\earth$ can be purely terrestrial (iron core plus silicate mantle). Observationally, the largest planet that is consistent with our terrestrial models is Kepler-20b, with $M=8.7\pm 2.2$ M$_\earth$ and $R=1.91_{-0.21}^{+0.12}$ R$_\earth$. In a transitional regime of $\sim$1.75-2.8 R$_\earth$ (which may correspond with an observed break in the planetary occurence function, \citealt{2013ApJS..204...24B}), the abundance of both water and H$_2$-He may be important, but the effect of the water fraction on the radius diminishes for larger planets.

Determining the composition and structure of solid planets from mass and radius observations is more ambiguous than for planets with gaseous envelopes due to significant degeneracies and uncertainties. There is overlap in radius with different iron fractions and water fractions over the mass range at which planets with potentially-solid compositions have been observed. Further research is needed to standardize equations of state for planetary models and reduce uncertainties.

There remains a degeneracy between composition and envelope entropy, and more detailed atmospheric and evolutionary models are needed to estimate the atmospheric entropy of exoplanets (to a precision of $\sim0.1 k_B/B$) in order to make more accurate determinations of their compositions and envelope/core mass partitions. On the observational side, more precise mass measurements are needed to better constrain core masses, as well as envelope masses in the case of smaller planets. More planet detections with overall better statistics are also needed to fully populate the mass-radius diagram and determine the distinct populations of planets (if any) in this regime.

\acknowledgements

A.B. would like to acknowledge support in part under NASA ATP grant NNX07AG80G, HST grants HST-GO-12181.04-A, HST-GO-12314.03-A, HST-GO-12473.06-A, and HST-GO-12550.02, and JPL/Spitzer Agreements 1417122, 1348668, 1371432, 1377197, and 1439064.


\appendix

\section{Radius Code Verification}
\label{radius_code_verification}

Our computational procedure for deriving planet structural profiles and the corresponding mass-radius relationships is as follows: We begin with a guess of central pressure and integrate the equations of hydrostatic equilibrium out until the pressure is zero. For differentiated planets, we dictate a boundary mass for the core material, at which we switch to the equation of state for the new material, maintaining pressure continuity. Our code admits an arbitrary number of layers of different materials. Because it is impossible to know the total mass or radius before the integration is performed, we use an iterative Newton-Raphson scheme to produce a planet of specified mass.

For hydrogen-helium envelopes, we assume an adiabatic pressure and density profile by setting a constant entropy, usually at 5.5$-$6.5 $k_B$ per baryon. This is a good approximation for convective envelopes, but not for radiative atmospheres, with the caveat that highly irradiated planets may be radiative to significant depths \citep{1996ApJ...459L..35G,2000ApJ...534L..97B}. The code also relies on extrapolations for pressures of less than 10 bars. (Pressures this low are not encountered in solid planets.)

We use a fourth-order Runge-Kutta scheme to solve the equations of hydrodynamic equilibrium:
\begin{eqnarray}
\frac{dP}{dr}=-\frac{Gm\rho}{r^2} \\
\frac{dm}{dr}=4\pi r^2\rho,
\end{eqnarray}
where \(r\) is the radius, \(\rho\) is the mass density, \(P(\rho)\) is the pressure given by the EOS, and \(m\) is the mass interior to \(r\). We use various radius step sizes ranging from 10$-$100m depending on the size of the planet model.

We test our code with the polytropic equation of state, \(P=K\rho^{1+\frac{1}{n}}\) with \(n=\)1.5, 2, 2.5, and 3. Our results for the constants \(\rho_c/\bar{\rho}\) (scaled density) and \(\xi\) (scaled radius) agree with the values found in \citet{1939isss.book.....C} to a part in \(10^5\).


\section{Equations of State}
\label{eos}

For solid planets, we implement two semi-empirical equations of state for most materials. The first is the third-order Birch-Murnagham EOS, given by
\begin{equation}
P_3 = \frac{3}{2}K_0 \left[x^{\frac{7}{3}}-x^{\frac{5}{3}}\right] \left\{1+\frac{3}{4}\left(K'_0-4\right)\left[x^{\frac{2}{3}}-1\right]\right\},
\end{equation}
where \(x\) is the ratio \(\rho / \rho_0\), \(\rho_0\) is the zero-pressure density, \(K_0\) is the bulk modulus at \(\rho=\rho_0\), and \(K'_0\) is the pressure derivative of the bulk modulus at \(\rho=\rho_0\). Values for the second pressure derivative of the bulk modulus at \(\rho=\rho_0\), \(K''_0\), are in most cases not available, but for materials which have a known \(K''_0\), a fourth-order term can be added to the third-order B-M EOS:
\begin{equation}
\begin{aligned}
P_4 = &P_3 + \frac{3}{2}K_0 \left[x^{\frac{7}{3}}-x^{\frac{5}{3}}\right]\frac{3}{8}\left(x^{\frac{2}{3}}-1\right)^2 \\
&\times \left[K_0 K''_0+K'_0\left(K'_0-7\right)+\frac{143}{9}\right].
\end{aligned}
\end{equation}
The second semi-empirical equation of state that we implement is the Vinet EOS, given by
\begin{equation}
P = 3 K_0 x^{2/3} \left[1-x^{-\frac{1}{3}}\right] \exp{\left(\frac{3}{2}\left(K'_0 - 1\right)\left[1-x^{-\frac{1}{3}}\right]\right)} .
\end{equation}
A summary of the equations of state used in our models is given in Table \ref{eos_ours_table}.

\begin{table}[tbhp]
\caption{Equations of State Used for this Project\(^{a}\)}
\begin{center}
\begin{tabular}{l|l|l|l|l|l}
\hline
Material                 & EOS   & \(\rho_0\) (g cm\(^{-3}\)) & K\(_0\) (Mbar)& K\('_0\) & Ref \\
\hline
Fe (\(\epsilon\))        & Vinet   &  8.267                     & 1.634         & 5.38     & 1   \\
MgSiO\(_3\)              & Vinet   &  4.064                     & 2.48          & 3.91     & 2   \\
(perovskite)             &         &                            &               &          &     \\
MgO                      & B-M3    &  3.5833                    & 1.602         & 3.99     & 3   \\
(periclase)              &         &                            &               &          &     \\
SiC (ZB)\(^b\)           & B-M3    &  3.350                     & 2.271         & 3.79     & 4   \\
SiC (RS)\(^b\)           & B-M3    &  4.256                     & 2.666         & 4.64     & 4   \\
Diamond                  & Vinet   &  3.5171                    & 4.45          & 4.0      & 5   \\
Platinum                 & Vinet   & 21.46                      & 2.70          & 5.64     & 6   \\
H\(_2\)O Ice VII         & Vinet   &  1.4876                    & 1.49          & 6.2      & 7   \\
H$_2$-He                 & Tabular &                            &               &          & 8   \\
\hline
\end{tabular}
\end{center}
\(^a\)All of the values are for materials at zero pressure and temperature. \(\rho_0\) is the density, K\(_0\) is the bulk modulus, and K\('_0\) is the pressure derivative of the bulk modulus.

\(^b\)The SiC EOS with zincblende (ZB) structure is used for pressures up to 0.75 Mbar, beyond which the EOS with rock salt (RS) structure is used.
\tablerefs{
(1) \citet{2006PhRvL..97u5504D};
(2) \citet{2004EaPSL.224..241T};
(3) \citet{2001JGR...106..515S};
(4) \citet{2008PhyB..403.3543L};
(5) \citet{2003PhRvB..68i4107K};
(6) \citet{2008PhRvB..78b4304S};
(7) \citet{1997PhRvB..56.5781W};
(8) \citet{1995ApJS...99..713S};
}
\label{eos_ours_table}
\end{table}

\begin{figure}[htp]
\includegraphics[width=\textwidth]{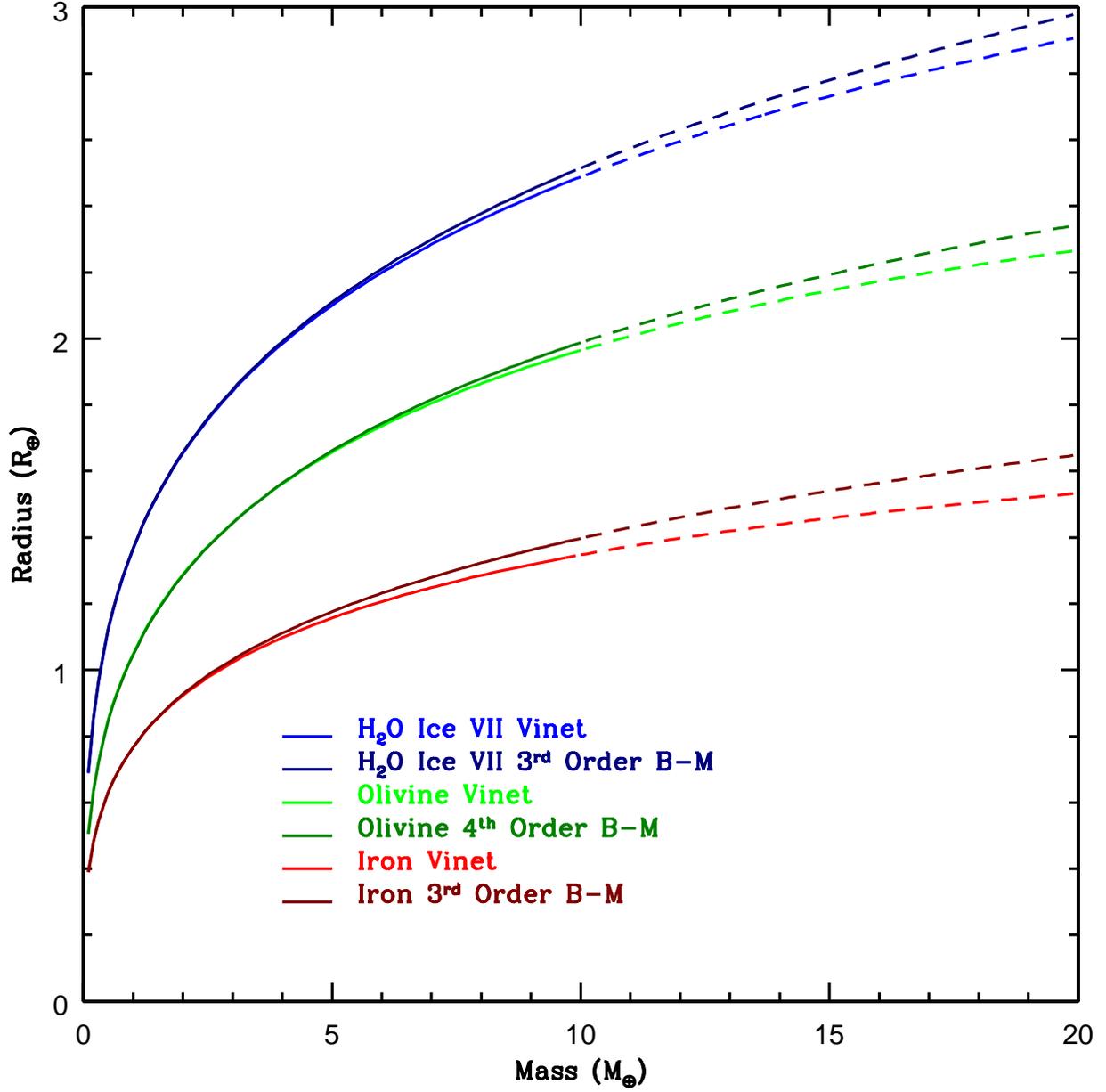}
\caption{Comparison of mass-radius curves computed with Vinet and Birch-Murnaghan EOSs for water (in the form of Ice VII), Mg$_2$SiO$_4$ (olivine), and iron. The two EOS fits agree in the low-pressure limit where the uncompressed density dominates, but deviate at high pressures. For context, the EOS fits for olivine differ by only 0.2\% for 1 Earth-mass planets, but differ increasingly as planet mass increases (to 6.8\% for 20 Earth-mass planets).}
\label{eos_fit_comparison}
\end{figure}

Which EOS fit is used for a given material makes little difference at low pressures where the zero-pressure density dominates. In Figure \ref{eos_fit_comparison}, models using the Vinet and B-M semi-empirical equations of state for pure MgSiO\(_3\) (perovskite) are used to generate mass-radius curves, which differ by only 0.3\% for 1 Earth-mass planets, 0.3\% for 5 Earth-mass planets, and 1.2\% for 10 Earth-mass planets. These results agree with those reported in \citet{2007ApJ...669.1279S}, who found that the mass-radius curves for low-mass exoplanets depend only on the uncompressed density. The differences become more significant for planets of higher mass (and, therefore, higher internal pressure), amounting to \(\sim\)3.3\% for 20 Earth-mass silicate planets and 7.0\% for 20 Earth-mass iron planets, although we do not expect to see solid planets of this size without H$_2$-He envelopes \citep{2013arXiv1311.0329L,2014ApJ...783L...6W}. A more detailed summary of differences is given in Table \ref{eos_compare}.

\begin{table}[tbhp]
\caption{Difference in Radius Between Planets with Vinet and B-M Equations of State}
\begin{center}
\begin{tabular}{l|l|l|l}
\hline
Mass (M$_\earth$) & Pure Fe & Pure MgSiO$_3$ & Pure H$_2$O \\
\hline
1.0               & 0.072\% & -0.286\%        & 0.239\%    \\
5.0               & 1.64\%  & 0.251\%         & 0.417\%    \\
10.0              & 3.58\%  & 1.17\%          & 1.06\%     \\
20.0              & 7.02\%  & 3.27\%          & 2.38\%     \\
\hline
\end{tabular}
\end{center}
\label{eos_compare}
\end{table}

\begin{figure}[htp]
\includegraphics[bb = -120 167 500 623,clip,width=\textwidth]{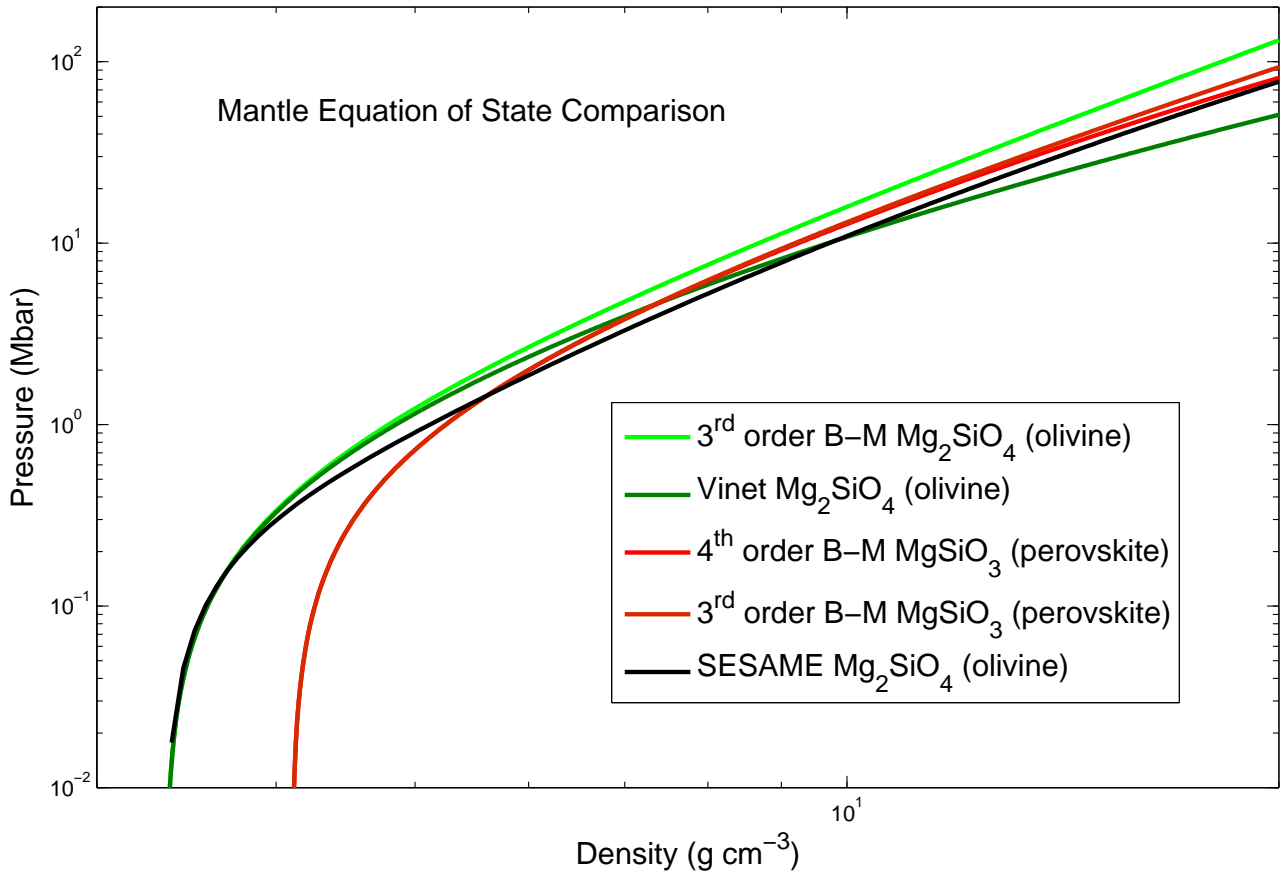}
\caption{Comparison of various equations of state for mantle materials. At high pressures (\(>\)1 Mbar) the equations of state for the two materials become similar, with the greatest differences between the various equations of state occurring for Mg\(_2\)SiO\(_4\) (olivine). This implies that for mantle materials at high pressures, the choice of material has a less significant effect than the choice of EOS fit. However, at pressures approaching zero, the uncompressed density dominates, making the choice of material far more significant than the choice of EOS fit at low pressures.}
\label{mantle_eos_comparison}
\end{figure}

To better illustrate the differences between the various equations of state, the pressure-density curves for various EOSs fits for Mg\(_2\)SiO\(_4\) (olivine) and MgSiO$_3$ (perovskite) are plotted in Figure \ref{mantle_eos_comparison} over the run of pressures found in super-Earths of mass up to 20 Earth masses (without gaseous envelopes). The curves show significant divergence in pressure at high densities, particularly for the different Mg$_2$SiO$_4$ EOSs, but remarkable similarities between the two materials, demonstrating that for rocky material, the choice of equation of state is more important than the precise chemical composition. For this reason, we do not address phase transitions in the mantle. This similarity breaks down below $\sim$1 Mbar, however, where the uncompressed density dominates. Notably, the $<1$ Mbar range includes most of the mantle of a 1-$M_\earth$ Earth-like planet, so the choice of phase transitions and the choice of silicate material could have a significant effect for lower-mass planets.

For context, internal pressure and density profiles for model 1-M$_\earth$ planets of various compositions are plotted in Figure \ref{density_pressure_profiles}. \citet{2000ipei.book.....P} and \citet{1996JPCM....8...67H} report that the Vinet EOS is more suited to extrapolation to high pressures. However, as Table \ref{eos_author_table} demonstrates, a wide variety of EOS data and fits have been used in the literature. Unless otherwise specified, we use a Vinet EOS in building our models, but we also use a 3rd-Order B-M EOS in some cases, as indicated in Table \ref{eos_ours_table}. For H$_2$-He envelopes, we use the EOS of \citet{1995ApJS...99..713S}.

We do not employ thermal corrections in our model, but we can estimate their effect from the coefficients of thermal expansion. The volume thermal expansion coefficient at high pressures has been measured to be $2-3\times 10^{-5}$ K$^{-1}$ for iron \citep{1990JGR....9521731B} and $4\times 10^{-5}$ K$^{-1}$ for MgSiO$_3$ \citep{1986Natur.319..214K}. Taking $3\times 10^{-5}$ K$^{-1}$ as an average, or a linear coefficient of $1\times 10^{-5}$ K$^{-1}$, and an internal temperature of 5,000 K, typical of Earth's core \citep{1994CRASB.318..341P}, we can estimate a thermal correction to the planetary radius of about 5\%. However, the internal temperature may be significantly higher for larger planets.

Alternatively, we can find the total error produced by our code from an Earth-analog model. Specifically a 1-M$_\earth$ planet model with an ``Earth-like'' composition produced by our code has a radius of 0.972 R$_\earth$, an error of 2.8\%. This includes the error introduced by thermal expansion compared with our zero-temperature model, the errors in the equations of state, and the error introduced by the simplifying assumption of a two-layer model. Since the EOS-induced errors are small for a 1-M$_\earth$ planet, it is possible that thermal expansion is a significant contributor to this error.

We compare our mass-radius results for a simple iron core/MgSiO\(_3\)-mantle planet to those obtained by \citet{2007ApJ...669.1279S}, to determine the effect of any differences in the EOS. We test a pure MgSiO\(_3\) composition, a 32.5\% iron core mass fraction (CMF) (``Earth-like'') composition, a 70\% CMF (``Mercury-like'') composition, and a pure iron composition for masses ranging from \(0.5M_\earth\) to \(20M_\earth\). We find agreement to within 1\% in all cases, with the exception of pure iron planets, which differ by up to -2.23\% for $20 M_\earth$. This is likely due to the effect on the EOS of the higher central pressures of iron planets.
Planets of mass 10 M\(_\earth\) composed of pure iron, 70\% iron, 32.5\% iron, and pure MgSiO\(_3\) (perovskite) have central pressures of 88 Mbar, 78 Mbar, 54 Mbar, and 18 Mbar, respectively.

\begin{figure}[htp]
\includegraphics[width=\textwidth]{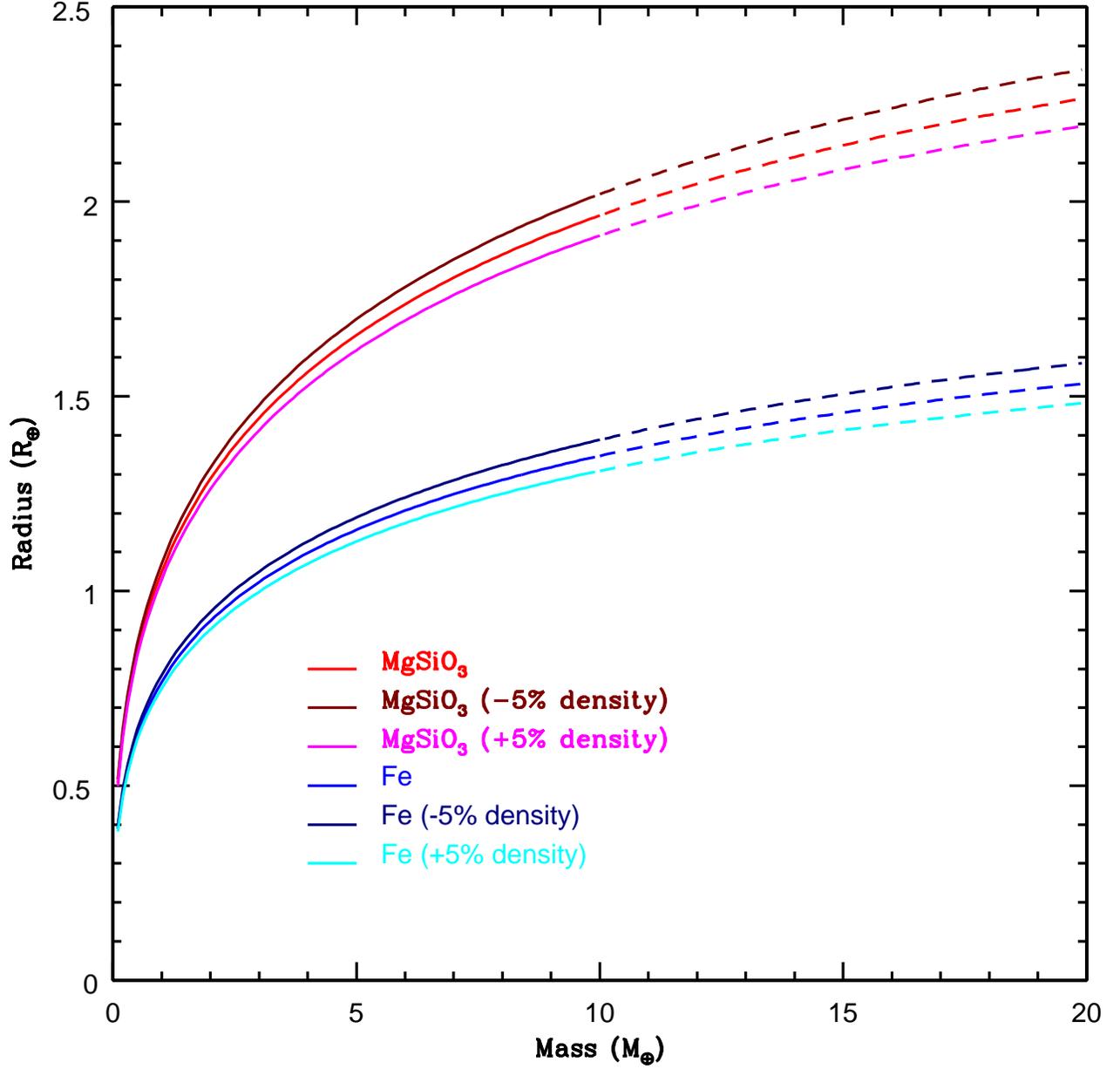}
\caption{Mass-radius curves for pure iron and silicate planets. A proportional EOS error in density is applied to investigate the effects of EOS ambiguities on terrestrial planet models. For a 10\% swing in density at a given pressure, we find changes in radius ranging from -3.5\% for a silicate, 0.1-M$_\earth$ planet to -6.4\% for an iron, 20-M$_\earth$ planet.}
\label{radius_vs_mass_eos_error1}
\end{figure}

As Figures \ref{eos_fit_comparison} and \ref{mantle_eos_comparison} demonstrate, there can be slight differences between different EOS fits. Ambiguities are especially prominent in the 1 to 1000 Mbar pressure range, the latter being above pressures one can probe with constant-temperature experiments, but below where the Thomas-Fermi-Dirac EOS becomes applicable \citep{1987Natur.330..737H}. In order to investigate the effect of EOS ambiguities on a terrestrial planet model, we multiply the density for all pressures by a constant error factor. The resulting mass-radius curves for pure iron and pure silicate planets are given in Figure \ref{radius_vs_mass_eos_error1}.

For a 10\% swing in density at a given pressure (on the order of the differences between EOSs in Figure \ref{mantle_eos_comparison}), we find changes in radius for pure MgSiO$_3$ planets of -3.5\% (0.1 M$_\earth$), -4.0\% (1.0 M$_\earth$), -4.6\% (5.0 M$_\earth$), -5.0\% (10.0 M$_\earth$), and -5.5\% (20.0 M$_\earth$). For pure iron planets, we find changes of -3.7\% (0.1 M$_\earth$), -4.4\% (1.0 M$_\earth$), -5.0\% (5.0 M$_\earth$), -5.3\% (10.0 M$_\earth$), and -6.4\% (20.0 M$_\earth$).

While most equations of state for a given material will agree on the low-pressure density (as it is easily measured), the differences between equations of state in the 1 to 1000 Mbar range could exceed 5\%. Because of this, the ambiguities in radius we report are relevant to high-mass planets (without envelopes) and solid cores whose central pressures lie where the equation of state is ambiguous. In particular, the uncertainties in radius due to the equation of state are similar in magnitude to the change in radius caused by differentiation and different mantle material (e.g. silicates vs. silicon carbide), so more accurate equations of state are needed to distinguish these cases.


\bibliographystyle{apj}
\bibliography{apj-jour,refs}


\clearpage

\clearpage


\pagebreak

\begin{table}[tbhp]
\caption{Earth-Like Core/H$_2$-H\MakeLowercase{e} Envelope Planets with $M_{env}=0.1$ M$_\earth$, $s=6.0\,k_B/B$}
\begin{center}
\begin{tabular}{l|l|l|l}
\hline
Core Mass      & Radius         & Central Pressure & Central Density \\
(M\(_\earth\)) & (R\(_\earth\)) &(Mbar)            & (g cm\(^{-3}\)) \\
\hline
0.4 & 4.26148 & 1.97466 & 12.5530 \\
0.6 & 3.47101 & 2.28146 & 13.5698 \\
0.8 & 3.06142 & 3.66167 & 14.4290 \\
1   & 2.79890 & 4.50292 & 15.1893 \\
2   & 2.33196 & 8.82066 & 18.2349 \\
3   & 2.21482 & 13.3997 & 20.6751 \\
4   & 2.17903 & 18.2692 & 22.8259 \\
5   & 2.17231 & 23.4375 & 24.8049 \\
6   & 2.17681 & 28.9078 & 26.6708 \\
7   & 2.18852 & 34.6839 & 28.4580 \\
8   & 2.20358 & 40.7688 & 30.1884 \\
9   & 2.21951 & 47.1661 & 31.8770 \\
10  & 2.23497 & 53.1950 & 33.5349 \\
12  & 2.26575 & 69.0382 & 36.7901 \\
14  & 2.29493 & 84.0288 & 40.0016 \\
16  & 2.32147 & 101.173 & 43.2015 \\
18  & 2.34494 & 119.786 & 46.4135 \\
20  & 2.36655 & 139.946 & 49.6562 \\
\hline
\end{tabular}
\end{center}
\label{0.1menv_6.0_table}
\end{table}

\begin{table}[tbhp]
\caption{Earth-Like Core/H$_2$-H\MakeLowercase{e} Envelope Planets with $M_c=2$ M$_\earth$, $s=5.5\,k_B/B$}
\begin{center}
\begin{tabular}{l|l|l|l}
\hline
Envelope Mass  & Radius         & Central Pressure & Central Density \\
(M\(_\earth\)) & (R\(_\earth\)) &(Mbar)            & (g cm\(^{-3}\)) \\
\hline
0.00 & 1.18851 & 8.69496 & 18.1594 \\
0.02 & 1.56409 & 8.73952 & 18.1862 \\
0.04 & 1.73518 & 8.76965 & 18.2044 \\
0.06 & 1.87084 & 8.79355 & 18.2187 \\
0.08 & 1.99033 & 8.81374 & 18.2308 \\
0.10 & 2.09529 & 8.83109 & 18.2412 \\
0.20 & 2.52114 & 8.89454 & 18.2790 \\
0.30 & 2.84728 & 8.93799 & 18.3049 \\
0.40 & 3.11830 & 8.97101 & 18.3245 \\
0.50 & 3.35079 & 8.99829 & 18.3407 \\
0.60 & 3.55310 & 9.02147 & 18.3544 \\
0.70 & 3.73529 & 9.04193 & 18.3665 \\
0.80 & 3.89788 & 9.06027 & 18.3773 \\
0.90 & 4.04563 & 9.07698 & 18.3872 \\
1.00 & 4.18264 & 9.09255 & 18.3963 \\
1.20 & 4.42436 & 9.12031 & 18.4127 \\
1.40 & 4.63533 & 9.14551 & 18.4275 \\
1.60 & 4.82146 & 9.16841 & 18.4409 \\
1.80 & 4.98358 & 9.18966 & 18.4534 \\
\hline
\end{tabular}
\end{center}
\label{2mc_5.5_table}
\end{table}

\begin{table}[tbhp]
\caption{Earth-Like Core/H$_2$-H\MakeLowercase{e} Envelope Planets with $M_c=2$ M$_\earth$, $s=6.0\,k_B/B$}
\begin{center}
\begin{tabular}{l|l|l|l}
\hline
Envelope Mass  & Radius         & Central Pressure & Central Density \\
(M\(_\earth\)) & (R\(_\earth\)) &(Mbar)            & (g cm\(^{-3}\)) \\
\hline
0.00 & 1.18851 & 8.69496 & 18.1594 \\
0.02 & 1.67729 & 8.73677 & 18.1846 \\
0.04 & 1.88652 & 8.76455 & 18.2013 \\
0.06 & 1.97443 & 8.78670 & 18.2146 \\
0.08 & 2.20039 & 8.80485 & 18.2255 \\
0.10 & 2.33232 & 8.82073 & 18.2350 \\
0.20 & 2.85687 & 8.87792 & 18.2691 \\
0.30 & 3.25957 & 8.91662 & 18.2922 \\
0.40 & 3.59448 & 8.94581 & 18.3095 \\
0.50 & 3.87978 & 8.97001 & 18.3239 \\
0.60 & 4.12971 & 8.99055 & 18.3361 \\
0.70 & 4.34681 & 9.00852 & 18.3467 \\
0.80 & 4.54684 & 9.04297 & 18.3565 \\
0.90 & 4.72604 & 9.03955 & 18.3651 \\
1.00 & 4.89030 & 9.05350 & 18.3733 \\
1.20 & 5.17685 & 9.07834 & 18.3880 \\
1.40 & 5.42263 & 9.10066 & 18.4011 \\
1.60 & 5.63461 & 9.12127 & 18.4132 \\
1.80 & 5.82015 & 9.14026 & 18.4244 \\
\hline
\end{tabular}
\end{center}
\label{2mc_6.0_table}
\end{table}

\begin{table}[tbhp]
\caption{Earth-Like Core/H$_2$-H\MakeLowercase{e} Envelope Planets with $M_c=2$ M$_\earth$, $s=6.5\,k_B/B$}
\begin{center}
\begin{tabular}{l|l|l|l}
\hline
Envelope Mass  & Radius         & Central Pressure & Central Density \\
(M\(_\earth\)) & (R\(_\earth\)) &(Mbar)            & (g cm\(^{-3}\)) \\
\hline
0.00 & 1.18851 & 8.69496 & 18.1594 \\
0.02 & 1.84538 & 8.73323 & 18.1825 \\
0.04 & 2.12095 & 8.75821 & 18.1975 \\
0.06 & 2.34125 & 8.77764 & 18.2092 \\
0.08 & 2.53227 & 8.79358 & 18.2187 \\
0.10 & 2.70473 & 8.80722 & 18.2269 \\
0.20 & 3.40678 & 8.85628 & 18.2562 \\
0.30 & 3.94117 & 8.88858 & 18.2755 \\
0.40 & 4.38415 & 8.91308 & 18.2901 \\
0.50 & 4.76993 & 8.93324 & 18.3021 \\
0.60 & 5.10259 & 8.95047 & 18.3123 \\
0.70 & 5.39890 & 8.96549 & 18.3212 \\
0.80 & 5.66089 & 8.97905 & 18.3293 \\
0.90 & 5.89601 & 8.99134 & 18.3366 \\
1.00 & 6.10857 & 9.00277 & 18.3422 \\
1.20 & 6.47809 & 9.02340 & 18.3555 \\
1.40 & 6.78834 & 9.04236 & 18.3667 \\
\hline
\end{tabular}
\end{center}
\label{2mc_6.5_table}
\end{table}

\begin{table}[tbhp]
\caption{Earth-Like Core/H$_2$-H\MakeLowercase{e} Envelope Planets with $M_c=5$ M$_\earth$, $s=6.0\,k_B/B$}
\begin{center}
\begin{tabular}{l|l|l|l}
\hline
Envelope Mass  & Radius         & Central Pressure & Central Density \\
(M\(_\earth\)) & (R\(_\earth\)) &(Mbar)            & (g cm\(^{-3}\)) \\
\hline
0.00 & 1.52092 & 23.2071 & 24.7228 \\
0.02 & 1.80841 & 23.2701 & 24.7445 \\
0.04 & 1.92515 & 23.3202 & 24.7626 \\
0.06 & 2.01793 & 23.3631 & 24.7781 \\
0.08 & 2.09911 & 23.4022 & 24.7922 \\
0.10 & 2.17720 & 23.4374 & 24.8049 \\
0.20 & 2.47107 & 23.5808 & 24.8565 \\
0.30 & 2.70888 & 23.6902 & 24.8957 \\
0.40 & 2.91421 & 23.7796 & 24.9277 \\
0.50 & 3.09611 & 23.8562 & 24.9550 \\
0.60 & 3.26045 & 23.9224 & 24.9786 \\
0.70 & 3.41082 & 23.9813 & 24.9996 \\
0.80 & 3.55018 & 24.0346 & 25.0186 \\
0.90 & 3.67956 & 24.0840 & 25.0361 \\
1.00 & 3.80286 & 24.1288 & 25.0520 \\
1.20 & 4.02328 & 24.2093 & 25.0806 \\
1.40 & 4.22312 & 24.2808 & 25.1059 \\
1.60 & 4.40458 & 24.3449 & 25.1285 \\
1.80 & 4.57070 & 24.4041 & 25.1495 \\
2.00 & 4.72406 & 24.4582 & 25.1685 \\
2.50 & 5.06239 & 24.5797 & 25.2113 \\
3.00 & 5.34830 & 24.6854 & 25.2484 \\
3.50 & 5.59620 & 24.7809 & 25.2819 \\
4.00 & 5.83033 & 24.8764 & 25.3153 \\
4.50 & 6.00929 & 24.9518 & 25.3416 \\
\hline
\end{tabular}
\end{center}
\label{5mc_6.0_table}
\end{table}

\begin{table}[tbhp]
\caption{Earth-Like Core/H$_2$-H\MakeLowercase{e} Envelope Planets with $M_c=10$ M$_\earth$, $s=6.0\,k_B/B$}
\begin{center}
\begin{tabular}{l|l|l|l}
\hline
Envelope Mass  & Radius         & Central Pressure & Central Density \\
(M\(_\earth\)) & (R\(_\earth\)) &(Mbar)            & (g cm\(^{-3}\)) \\
\hline
0.00 & 1.79615 & 53.5130 & 33.4467 \\
0.02 & 1.99228 & 53.6051 & 33.4688 \\
0.04 & 2.07032 & 53.6835 & 33.4876 \\
0.06 & 2.13342 & 53.7542 & 33.5046 \\
0.08 & 2.18642 & 53.8190 & 33.5201 \\
0.10 & 2.23596 & 53.8813 & 33.5350 \\
0.20 & 2.43488 & 54.1477 & 33.5987 \\
0.30 & 2.59648 & 54.3693 & 33.6515 \\
0.40 & 2.73667 & 54.5604 & 33.6971 \\
0.50 & 2.86268 & 54.7313 & 33.7377 \\
0.60 & 2.97828 & 54.8846 & 33.7741 \\
0.70 & 3.08500 & 55.0238 & 33.8071 \\
0.80 & 3.18559 & 55.1530 & 33.8377 \\
0.90 & 3.28040 & 55.2727 & 33.8660 \\
1.00 & 3.36982 & 55.3828 & 33.8920 \\
1.20 & 3.53525 & 55.5862 & 33.9400 \\
1.40 & 3.69025 & 55.7704 & 33.9834 \\
1.60 & 3.83086 & 55.9349 & 34.0222 \\
1.80 & 3.96291 & 56.0866 & 34.0578 \\
2.00 & 4.08644 & 56.2271 & 34.0908 \\
2.50 & 4.36723 & 56.5416 & 34.1644 \\
3.00 & 4.61361 & 56.8150 & 34.2283 \\
3.50 & 4.83412 & 57.0578 & 34.2850 \\
4.00 & 5.03420 & 57.2804 & 34.3368 \\
4.50 & 5.21535 & 57.4825 & 34.3837 \\
5.00 & 5.38451 & 57.6730 & 34.4279 \\
6.00 & 5.68357 & 58.0220 & 34.5087 \\
7.00 & 5.94431 & 58.3358 & 34.5811 \\
8.00 & 6.17595 & 58.6298 & 34.6488 \\
\hline
\end{tabular}
\end{center}
\label{10mc_6.0_table}
\end{table}

\begin{table*}[tbhp]
\caption{Pure MgSiO\(_3\) (perovskite) Planets}
\begin{center}
\begin{tabular}{l|l|l|l}
\hline
Mass           & Radius         & Central Pressure & Central Density \\
(M\(_\earth\)) & (R\(_\earth\)) &(Mbar)            & (g cm\(^{-3}\)) \\
\hline
0.2 & 0.634704 & 0.44969 & 4.66933 \\
0.4 & 0.791132 & 0.76609 & 5.00284 \\
0.6 & 0.897857 & 1.06129 & 5.27594 \\
0.8 & 0.980917 & 1.34800 & 5.51649 \\
1   & 1.04968  & 1.63112 & 5.73586 \\
2   & 1.28797  & 3.05157 & 6.66347 \\
3   & 1.44433  & 4.53165 & 7.45104 \\
4   & 1.56222  & 6.09528 & 8.17135 \\
5   & 1.65714  & 7.75175 & 8.85318 \\
6   & 1.73659  & 9.50646 & 9.51142 \\
7   & 1.80482  & 11.3635 & 10.1551 \\
8   & 1.86449  & 13.3270 & 10.7901 \\
9   & 1.91740  & 15.4003 & 11.4208 \\
10  & 1.96480  & 17.5872 & 12.0502 \\
12  & 2.04659  & 22.3176 & 13.3149 \\
14  & 2.11495  & 27.5538 & 14.5999 \\
16  & 2.17308  & 33.3339 & 15.9166 \\
18  & 2.22312  & 39.7017 & 17.2748 \\
20  & 2.26653  & 46.7077 & 18.6833 \\
\hline
\end{tabular}
\end{center}
\label{pv_planet_table}
\end{table*}

\begin{table*}[tbhp]
\caption{32.5\% Iron / 67.5\% MgSiO\(_3\) (perovskite) Differentiated Planets}
\begin{center}
\begin{tabular}{l|l|l|l}
\hline
Mass           & Radius         & Central Pressure & Central Density \\
(M\(_\earth\)) & (R\(_\earth\)) &(Mbar)            & (g cm\(^{-3}\)) \\
\hline
0.2 & 0.592384 & 1.08997 & 11.2072 \\
0.4 & 0.736224 & 1.94664 & 12.5157 \\
0.6 & 0.833926 & 2.77697 & 13.5207 \\
0.8 & 0.909741 & 3.60223 & 14.3719 \\
1.0 & 0.972372 & 4.43009 & 15.1266 \\
2   & 1.18850  & 8.69494 & 18.1594 \\
3   & 1.32962  & 13.2340 & 20.5953 \\
4   & 1.43571  & 18.0696 & 22.7440 \\
5   & 1.52093  & 23.2070 & 24.7217 \\
6   & 1.59214  & 28.6485 & 26.5866 \\
7   & 1.65321  & 34.3970 & 28.3729 \\
8   & 1.70657  & 40.4541 & 30.1021 \\
9   & 1.75383  & 46.8249 & 31.7898 \\
10  & 1.79616  & 53.5130 & 33.4467 \\
12  & 1.86914  & 67.8618 & 36.6994 \\
14  & 1.93012  & 83.5511 & 39.9084 \\
16  & 1.98200  & 100.637 & 43.1054 \\
18  & 2.02670  & 119.191 & 46.3144 \\
20  & 2.06553  & 139.285 & 49.5533 \\
\hline
\end{tabular}
\end{center}
\label{0.325Fe_planet_table}
\end{table*}

\begin{table*}[tbhp]
\caption{32.5\% Iron / 67.5\% MgSiO\(_3\) (perovskite) Undifferentiated Planets}
\begin{center}
\begin{tabular}{l|l|l|l}
\hline
Mass           & Radius         & Central Pressure & Central Density \\
(M\(_\earth\)) & (R\(_\earth\)) &(Mbar)            & (g cm\(^{-3}\)) \\
\hline
0.2 & 0.593662 & 0.596268 & 5.82786 \\
0.4 & 0.737793 & 1.03360  & 6.33721 \\
0.6 & 0.835511 & 1.44955  & 6.75061 \\
0.8 & 0.911201 & 1.85902  & 7.11316 \\
1   & 0.973627 & 2.26762  & 7.44304 \\
2   & 1.18822  & 4.36307  & 8.83494 \\
3   & 1.32751  & 6.60057  & 10.0176 \\
4   & 1.43167  & 9.00490  & 11.1022 \\
5   & 1.51499  & 11.5858  & 12.1321 \\
6   & 1.58431  & 14.3501  & 13.1295 \\
7   & 1.64351  & 17.3037  & 14.1081 \\
8   & 1.69501  & 20.4523  & 15.0765 \\
9   & 1.74045  & 23.8027  & 16.0411 \\
10  & 1.78097  & 27.3617  & 17.0068 \\
12  & 1.85037  & 35.1346  & 18.9558 \\
14  & 1.90784  & 43.8386  & 20.9471 \\
16  & 1.95624  & 53.5474  & 22.9986 \\
18  & 1.99745  & 64.3474  & 25.1257 \\
20  & 2.03281  & 76.3381  & 27.3427 \\
\hline
\end{tabular}
\end{center}
\label{0.325Fe_planet_table_homogeneous}
\end{table*}

\begin{table*}[tbhp]
\caption{70\% Iron / 30\% MgSiO\(_3\) (perovskite) Differentiated Planets}
\begin{center}
\begin{tabular}{l|l|l|l}
\hline
Mass           & Radius         & Central Pressure & Central Density \\
(M\(_\earth\)) & (R\(_\earth\)) &(Mbar)            & (g cm\(^{-3}\)) \\
\hline
0.2 & 0.535687 & 1.39506 & 11.7182 \\
0.4 & 0.662772 & 2.54306 & 13.2557 \\
0.6 & 0.748595 & 3.67582 & 14.4425 \\
0.8 & 0.814977 & 4.81495 & 15.4518 \\
1   & 0.869685 & 5.96775 & 16.3500 \\
2   & 1.05786  & 12.0087 & 19.9881 \\
3   & 1.18039  & 18.5516 & 22.9409 \\
4   & 1.27243  & 25.6011 & 25.5667 \\
5   & 1.34640  & 33.1514 & 27.9993 \\
6   & 1.40823  & 41.2004 & 30.3062 \\
7   & 1.46131  & 49.7471 & 32.5268 \\
8   & 1.50776  & 58.7933 & 34.6864 \\
9   & 1.54895  & 68.3429 & 36.8026 \\
10  & 1.58589  & 78.4005 & 38.8879 \\
12  & 1.64978  & 100.070 & 43.0034 \\
14  & 1.70339  & 123.870 & 47.0883 \\
16  & 1.74922  & 149.882 & 51.1807 \\
18  & 1.78893  & 178.211 & 55.3085 \\
20  & 1.82366  & 208.966 & 59.4937 \\
\hline
\end{tabular}
\end{center}
\label{0.7Fe_planet_table}
\end{table*}

\begin{table*}[tbhp]
\caption{70\% Iron / 30\% MgSiO\(_3\) (perovskite) Undifferentiated Planets}
\begin{center}
\begin{tabular}{l|l|l|l}
\hline
Mass           & Radius         & Central Pressure & Central Density \\
(M\(_\earth\)) & (R\(_\earth\)) &(Mbar)            & (g cm\(^{-3}\)) \\
\hline
0.2 & 0.537303 & 0.914989 & 8.15393 \\
0.4 & 0.664291 & 1.63148  & 9.06921 \\
0.6 & 0.749565 & 2.33087  & 9.79947 \\
0.8 & 0.815181 & 3.03122  & 10.4338 \\
1   & 0.869019 & 3.73887  & 11.0074 \\
2   & 1.05228  & 7.45597  & 13.4028 \\
3   & 1.16984  & 11.5200  & 15.4184 \\
4   & 1.25710  & 15.9491  & 17.2568 \\
5   & 1.32649  & 20.7482  & 18.9951 \\
6   & 1.38396  & 25.9242  & 20.6731 \\
7   & 1.43288  & 31.4829  & 22.3140 \\
8   & 1.47531  & 37.4317  & 23.9330 \\
9   & 1.51264  & 43.7801  & 25.5410 \\
10  & 1.54585  & 50.5397  & 27.1464 \\
12  & 1.60261  & 65.3367  & 30.3725 \\
14  & 1.64946  & 81.9292  & 33.6493 \\
16  & 1.68884  & 100.440  & 37.0047 \\
18  & 1.72233  & 121.008  & 40.4615 \\
20  & 1.75106  & 143.796  & 44.0398 \\
\hline
\end{tabular}
\end{center}
\label{0.7Fe_planet_table_homogeneous}
\end{table*}

\begin{table*}[tbhp]
\caption{Pure Iron Planets}
\begin{center}
\begin{tabular}{l|l|l|l}
\hline
Mass           & Radius         & Central Pressure & Central Density \\
(M\(_\earth\)) & (R\(_\earth\)) &(Mbar)            & (g cm\(^{-3}\)) \\
\hline
0.2 & 0.481284 & 1.47742 & 11.8462 \\
0.4 & 0.591383 & 2.71052 & 13.4467 \\
0.6 & 0.664753 & 3.93539 & 14.6857 \\
0.8 & 0.720967 & 5.17286 & 15.7419 \\
1   & 0.766975 & 6.42967 & 16.6836 \\
2   & 0.923092 & 13.0654 & 20.5136 \\
3   & 1.02317  & 20.3131 & 23.6389 \\
4   & 1.09763  & 28.1668 & 26.4291 \\
5   & 1.15705  & 36.6171 & 29.0228 \\
6   & 1.20647  & 45.6585 & 31.4895 \\
7   & 1.24872  & 55.2906 & 33.8702 \\
8   & 1.28555  & 65.5144 & 36.1910 \\
9   & 1.31813  & 76.3373 & 38.4707 \\
10  & 1.34727  & 87.7620 & 40.7217 \\
12  & 1.39751  & 112.461 & 45.1776 \\
14  & 1.43952  & 139.698 & 49.6177 \\
16  & 1.47532  & 169.583 & 54.0822 \\
18  & 1.50626  & 202.237 & 58.6007 \\
20  & 1.53329  & 237.809 & 63.1980 \\
\hline
\end{tabular}
\end{center}
\label{Fe_planet_table}
\end{table*}

\clearpage

\end{document}